\documentclass[twocolumn,twoside]{IEEEtran}
\usepackage{mathpple}
\usepackage{times}
\usepackage{amsmath}  
\usepackage{amssymb}  
\usepackage{mathrsfs} 
\usepackage{theorem}  
\usepackage{cite}     
\usepackage{comment}  
\usepackage{upref}
\usepackage{amsfonts}
\usepackage{verbatim}
\usepackage[dvipsnames,usenames]{color}
\usepackage{enumerate}
\usepackage{cuted}
\usepackage[shortlabels]{enumitem}
\usepackage{dsfont}

\usepackage{algorithm}
\usepackage{times,color}
\usepackage{graphicx}
\usepackage{float}
\usepackage{color}
\usepackage{mathtools}
\usepackage{graphicx}
\usepackage{listings}
\usepackage{tikz}
\usepackage{multicol}
\usepackage{psfrag}

\usepackage{pgfplots}
\usepackage{hhline}
\usepackage{xcolor,colortbl}
\usepackage{algorithm}
\usepackage[noend]{algpseudocode}

\usepackage{wrapfig}
\usepackage{subfigure}
\usepackage[top=0.58in, bottom=0.58in, left=0.6in, right=0.6in]{geometry}
\usepackage{kantlipsum,cuted}
\usepackage[all]{xy}
\usepackage{algorithmicx}
\algrenewcommand\alglinenumber[1]{\scriptsize #1:}
\algrenewcommand\algorithmicindent{1em}%
\usepackage{latexsym}
\usepackage{multirow}

\usepackage{tabu}
\usepackage{lipsum}
\usepackage{blkarray, bigstrut}
\usepackage{comment}

\newcommand{\field}[1]{\mathbb{#1}}

\newcommand{\F}{\field{F}}

\newcommand{\RN}[1]{%
	\textup{\uppercase\expandafter{\romannumeral#1}}%
}

\newtheorem{innercustomgeneric}{\customgenericname}
\providecommand{\customgenericname}{}
\newcommand{\newcustomtheorem}[2]{%
	\newenvironment{#1}[1]
	{%
		\renewcommand\customgenericname{#2}%
		\renewcommand\theinnercustomgeneric{##1}%
		\innercustomgeneric
	}
	{\endinnercustomgeneric}
}
\newcustomtheorem{customproposition}{Proposition}
\newcustomtheorem{customtheorem}{Theorem}
\newcustomtheorem{customlemma}{Lemma}
\newcustomtheorem{customclaim}{Claim}
\newcustomtheorem{customconjecture}{Conjecture}

\makeatletter
\def\BState{\State\hskip-\ALG@thistlm}

\makeatother

\algnewcommand{\Initialize}[1]{%
	\State \textbf{Initialize:}
	\Statex \hspace*{\algorithmicindent}\parbox[t]{.8\linewidth}{\raggedright #1}
}

\newcommand{\be}[1]{\begin{equation}\label{#1}}
\newcommand{\ee}{\end{equation}}

\newcommand{\bc}{\begin{center}}
\newcommand{\ec}{\end{center}}

\newcommand{\cB}{{\cal B}}

\newcommand{\cF}{{\cal F}}
\newcommand{\cG}{{\cal G}}
\newcommand{\cH}{{\cal H}}

\newcommand{\cM}{{\cal M}}
\newcommand{\cN}{{\cal N}}
\newcommand{\cO}{{\cal O}}
\newcommand{\cP}{{\cal P}}

\newcommand{\cR}{{\cal R}}
\newcommand{\cS}{{\cal S}}

\newcommand{\bfa}{{\boldsymbol a}}

\newcommand{\bfe}{{\boldsymbol e}}

\newcommand{\bfg}{{\boldsymbol g}}

\newcommand{\bfu}{{\boldsymbol u}}
\newcommand{\bfv}{{\boldsymbol v}}
\newcommand{\bfw}{{\boldsymbol w}}
\newcommand{\bfx}{{\boldsymbol x}}
\newcommand{\bfy}{{\boldsymbol y}}

\newcommand{\bfC}{{\mathbf C}}

\newcommand{\bfG}{{\mathbf G}}

\renewcommand{\leq}{\leqslant}

\renewcommand{\geq}{\geqslant}

\newcommand{\Cref}[1]{Co\-rol\-la\-ry\,\ref{#1}}

\theoremstyle{plain} \theorembodyfont{\normalfont\slshape}

\newtheorem{thm}{Theorem$\!$}
\newenvironment{theorem}{\begin{thm}\hspace*{-1ex}{\bf.}}{\end{thm}}

\newtheorem{prop}[thm]{Proposition$\!$}

\newtheorem{lem}[thm]{Lemma$\!$}
\newenvironment{lemma}{\begin{lem}\hspace*{-1ex}{\bf.}}{\end{lem}}

\newtheorem{cor}[thm]{Corollary$\!$}
\newenvironment{corollary}{\begin{cor}\hspace*{-1ex}{\bf.}}{\end{cor}}

\newtheorem{cons}[thm]{Construction$\!$}

\newtheorem{defi}[thm]{Definition$\!$}
\newenvironment{definition}{\begin{defi}\hspace*{-1ex}{\bf.}}{\end{defi}}

\newtheorem{cl}{Claim}
\newenvironment{claim}{\begin{cl}\hspace*{-1ex}{\bf .}}{\end{cl}}
\theorembodyfont{\normalfont}

\newtheorem{conj}{Conjecture$\!$}
\newenvironment{conjecture}{\begin{conj}\hspace*{-1ex}{\bf .}}{\end{conj}}

\newtheorem{exam}{Example$\!$}
\newenvironment{example}{\begin{exam}\hspace*{-1ex}{\bf .}}{\end{exam}}

\newtheorem{remrk}{Remark$\!$}

\definecolor{Codecolor}{named}{White}  

\newcommand{\Copen}{\mbox{\{\kern-5.50pt\{}}
\newcommand{\Cclose}{\mbox{\}\kern-5.50pt\}}}
\newcommand{\Cslash}{\mbox{$\backslash\kern-6.02pt\backslash$}}

\newcommand{\bs}{\textmd{-}}

\newcommand{\e}[1]{\textcolor{red}{#1}}

\newcommand{\HG}{$HG$-matrix}
\newcommand{\HGG}{$HG$-matrix }
\newcommand{\HGs}{$HG$-matrices }

\begin{document}
\title{Almost Optimal Construction of Functional Batch Codes Using Hadamard Codes}
\author{\large Lev~Yohananov,~\IEEEmembership{Student Member,~IEEE},  and Eitan~Yaakobi,~\IEEEmembership{Senior Member,~IEEE} 
\thanks{L. Yohananov, and E. Yaakobi are with the Department of Computer Science, Technion --- Israel Institute of Technology, Haifa 3200003, Israel (e-mail: \texttt{\{levyohananov,yaakobi\}@cs.technion.ac.il}).}
\vspace{-3ex}}
\maketitle	
\begin{abstract}
	A \textit{functional $k$-batch} code of dimension $s$ consists of $n$ servers
	storing linear combinations of $s$ linearly independent information bits. Any multiset request of size $k$ of linear combinations (or requests) of the information bits can be recovered by $k$ disjoint subsets of the servers. The goal under this paradigm is to find the minimum number of servers for given values of $s$ and $k$. A recent conjecture states that for any $k=2^{s-1}$ requests the optimal solution requires $2^s-1$ servers. This conjecture is verified for $s\leq 5$ but previous work could only show that codes with $n=2^s-1$ servers can support a solution for $k=2^{s-2} + 2^{s-4} + \left\lfloor \frac{ 2^{s/2}}{\sqrt{24}} \right\rfloor$ requests. This paper reduces this gap and shows the existence of codes for $k=\lfloor \frac{5}{6}2^{s-1} \rfloor - s$ requests with the same number of servers. 
	Another construction in the paper provides a code with $n=2^{s+1}-2$ servers and $k=2^{s}$ requests, which is an optimal result.
	These constructions are mainly based on Hadamard codes and equivalently provide constructions for \textit{parallel Random I/O (RIO)} codes.
\end{abstract}


\section{Introduction}
Motivated by several applications for load-balancing in storage and cryptographic protocols, \textit{batch codes} were first proposed by Ishai et al.~\cite{Ishai}. A batch code encodes a length-$s$ string $\bfx$ into $n$ strings, where each string corresponds to a \textit{server}, such that each batch request of $k$ different bits (and more generally symbols) from $\bfx$ can be decoded by reading at most $t$ bits from every server. This decoding process corresponds to the case of a single-user. There is an extended variant for batch codes~\cite{Ishai} which is intended for a multi-user application instead of a single-user setting, known as the \textit{multiset batch codes}. Such codes have $k$ different users and each requests a single data item. Thus, the $k$ requests can be represented as a \emph{multiset} of the bits since the requests of different users may be the same, and each server can be accessed by at most one user. 

A special case of multiset batch codes, referred as \emph{primitive batch codes}, is when each server contains only one bit. 
The goal of this model is to find, for given $s$ and $k$, the smallest $n$ such that a primitive batch code exists. This problem was considered in several papers; see e.g.~\cite{Asi,Buzaglo,Ishai,Rawat,Vardy}. By setting the requests to be a multiset of linear combinations of the $s$ information bits, a batch code is generalized into a \textit{functional batch code}~\cite{Zhang}. Again,  given $s$ and $k$, the goal is to find the smallest $n$ for which a functional $k$-batch code exists.

Mathematically speaking, an $FB\bs(n,s,k)$ functional $k$-batch  code (and in short $FB\bs(n,s,k)$ code) of dimension $s$ consists of $n$ servers storing linear combinations of $s$ linearly independent information bits. Any multiset of size $k$ of linear combinations from the linearly independent information bits, can be recovered by $k$ disjoint subsets of servers. If all the $k$ linear combinations are the same, then the servers form an $FP\bs(n,s,k)$  \textit{functional $k$-Private Information Retrieval (PIR)} code (and in short $FP\bs(n,s,k)$ code). 
Clearly, an $FP\bs(n,s,k)$ code is a special case of an $FB\bs(n,s,k)$ code. It was shown that functional $k$-batch codes  are equivalent to the so-called linear \textit{parallel random I/O (RIO) codes}, where RIO codes were introduced by Sharon and Alrod~\cite{Sharon}, and their parallel variation was studied in~\cite{Sun,Vajha}. Therefore, all the results for functional $k$-batch codes of this paper hold also for parallel RIO codes.  {If all the $k$ linear combinations are of a single information bit (rather than linear combinations of information bits), then the servers form an $B\bs(n,s,k)$  \textit{$k$-batch} code (and in short $B\bs(n,s,k)$ code). }

	 {The value $FP(s, k),B(s, k),FB(s, k)$ is defined to be the minimum number of servers required for the existence of an $FP\bs(n,s,k),$ $B\bs(n,s,k),FB\bs(n,s,k)$ code, respectively. Several upper and lower bounds can be found in~\cite{Zhang} on these values.
	Wang et al.~\cite{Wang} showed that for $k=2^{s-1}$, the length of an optimal $k$-batch code is $2^s-1$, that is, $B(s, k=2^{s-1}) = 2^s-1$. They also showed a {recursive decoding algorithm.}
	It was conjectured in~\cite{Zhang} that for the same value of $k$, the length of an optimal functional batch code is $2^s-1$, that is, $FB(s, k=2^{s-1}) = 2^s-1$.} Indeed, in~\cite{Yamawaki} this conjecture was proven for $s=3,4$, and in~\cite{Zhang}, by using a computer search, it was verified also for $s=5$.  However, the best-known result for $s>5$ only provides a construction of $FB\bs(2^s-1,s, 2^{s-2} + 2^{s-4} + \left\lfloor \frac{ 2^{s/2}}{\sqrt{24}} \right\rfloor)$ codes~\cite{Zhang}. This paper significantly improves this result and reduces the gap between the conjecture statement and the best-known construction. In particular, a construction of $FB\bs(2^s-1,s,\lfloor\frac{5}{6}\cdot 2^{s-1} \rfloor-s)$ codes is given. 
	To obtain this important result, we first show an existence of $FB\bs(2^s-1,s,\lfloor\frac{3}{4}\cdot 2^{s-1} \rfloor)$ code.
	Moreover, we show how to construct  $FB\bs(2^s +\lceil(3\alpha-2)\cdot2^{s-2}\rceil -1 ,s,\lfloor\alpha\cdot 2^{s-1}\rfloor)$ codes for all $2/3 \leq \alpha \leq 1$. Another result that can be found in~\cite{Zhang} states that $FP(s, 2^{s}) \leq 2^{s+1}-2$. In this case, the lower bound is the same, i.e., this result is optimal, see~\cite{Fazeli}. In this paper we will show that this optimality holds not only for functional PIR codes but also for the more challenging case of functional batch codes, that is, $FB(s, 2^{s}) =   2^{s+1}-2$.
	 {Lastly, we show a non-recursive  {decoding algorithm} for $B\bs(2^s,s,k=2^{s-1})$ codes. In fact, this construction holds not only for $k$ single bit requests (with respect to $k$-batch codes) but also for $k$ linear combinations of requests under some constraint that will be explained in the paper.}
	All the results in the paper are achieved using a generator matrix $G$ of a Hadamard codes~\cite{Arora} of length $2^s$ and dimension $s$, where the matrix's columns correspond to the servers of the $FB\bs(n,s,k)$ code.

The rest of the paper is organized as follows. In Section~\ref{sec:defs}, we formally define functional $k$-batch codes and summarize the main results of the paper. In Section~\ref{sec:cons1}, we show a construction of  $FB\bs(2^s +\lceil (3\alpha-2)\cdot2^{s-2}\rceil -1 ,s,\lfloor\alpha\cdot 2^{s-1}\rfloor)$ for $\alpha = 2/3$. This result is extended for all $2/3 \leq \alpha \leq 1$ in Section~\ref{sec:cons3}. In Section~\ref{sec:cons2}, a construction of $FB\bs(2^{s+1}-2,s,2^s)$  is presented.
  {In Section~\ref{sec:cons5}, we present our main result, i.e., a construction of $FB\bs(2^s-1,s,\lfloor\frac{5}{6}\cdot 2^{s-1} \rfloor-s)$  codes.
In Section~\ref{sec:extension1}  a construction of $B\bs(2^{s}-1,s,2^{s-1})$  is presented.}   Finally, Section~\ref{sec:conc} concludes the paper.

\section{Definitions}\label{sec:defs}
 
For a positive integer $n$ define $[n] = \{0,1,\dots,n-1\}$. All vectors and matrices in the paper are over $\F_2$.  
We follow the definition of functional batch codes as it was first defined in~\cite{Zhang}.
\begin{definition}\label{func_batch}
A \textbf{functional $k$-batch code} of length $n$ and dimension $s$ consists of $n$ servers
and $s$ information bits $x_0,x_1,\dots,x_{s-1}$. Each server stores a nontrivial linear combination
of the information bits (which are the coded bits), i.e.,  for all $j\in[n]$, the $j$-th server stores a linear
combination
\begin{align*}
&y_j =  x_{i_0}+x_{i_1}+\dots+x_{i_{\ell-1}},
\end{align*}
such that  $i_0,i_1,\dots,i_{\ell-1} \in [s]$.
For any request of $k$ linear bit combinations $v_0,v_1,\dots,v_{k-1}$ (not necessarily distinct) of the information bits, there are $k$ pairwise disjoint subsets $R_0,R_1,\dots,R_{k-1}$
of $[n]$ such that the sum of the linear combinations in the related servers of $R_i$, $i\in[k]$, is
$v_i$, i.e., 
$$ \sum_{j\in R_i}y_j =  v_i.$$ 
Each such $v_i$ will be called a \textbf{requested bit}  and each such subset $R_i$ will be called a  \textbf{recovery set}.

A functional $k$-batch code can be also represented by a \textit{linear code} with an $s \times n$ \textit{generator matrix}
$$G = [\bfg_0,\bfg_1,\dots,\bfg_{n-1}]$$ 
in $\F^{s\times n}_2$ in which the vector $\bfg_j$ has ones in positions 
$i_0,i_1,\dots,i_{\ell-1}$ if and only if the $j$-th server stores the linear
combination $x_{i_0}+x_{i_1}+\dots+x_{i_{\ell-1}}$. 
Using this matrix representation, a functional $k$-batch code is
an $s\times n$ generator matrix $G$, such that for any $k$ request vectors $\bfv_0,\bfv_1,\dots,\bfv_{k-1}\in\F^s_2$ (not necessarily distinct),
there are $k$ pairwise disjoint subsets of columns in $G$, denoted by $R_0,R_1,\dots,R_{k-1}$, such that the sum of the column vectors whose indices are in $R_j$ is equal to the \textbf{request vector} $\bfv_j$.  The set of all recovery sets $R_i,i\in[k]$, is called a \textbf{solution} for the $k$ request vectors.   {The sum of the column vectors whose indices are in $R_j$ will be called the \textbf{recovery sum}}.

A functional $k$-batch code of length $n$ and dimension $s$ over $\F^s_2$ is denoted by $FB\bs(n,s,k)$.  Every  request of $k$  vectors will be stored as columns in a matrix $M$ which is called the \textbf{request matrix} or simply the \textbf{request}.

\end{definition}

 {A \textbf{$k$-batch code} of length $n$ and dimension $s$ over $\F^s_2$, is denoted by $B\bs(n,s,k)$ and is defined similarly to functional $k$-batch codes as in Definition~\ref{func_batch} except of the fact that each request vector $\bfv_j\in\F^s_2$ is a unit vector.}
A \textbf{functional $k$-PIR code}~\cite{Zhang} of length $n$ and dimension $s$, denoted by $FP\bs(n,s,k)$,  is a special case of  $FB\bs(n,s,k)$ in which all the request vectors are identical. 
We first show some preliminary results on the parameters of  $FB\bs(n,s,k)$ and  $FP\bs(n,s,k)$ codes which are relevant to our work. For that, another definition is presented.

\begin{definition}
	Denote by $FB(s, k ),B(s,k),FP(s, k )$ the minimum length $n$ of any $FB\bs(n,s,k),B\bs(n,s,k),FP\bs(n,s,k)$ code, respectively.
\end{definition}

Most of the following results on $FB(s, k ),B(s,k)$ and $FP(s, k )$ can be found in~\cite{Zhang}, while the result in $(c)$ was verified for $s=3,4$ in~\cite{Yamawaki}.
\begin{theorem}\label{theo:0} 
For positive integers $s$ and $t$, the following properties hold:
	\begin{enumerate}[(a)]
		\item $FP(s, 2^{s-1}) = 2^s-1$.
		\item  $FP(st, 2^s) \leq 2t(2^s-1)$.
		\item For $s\leq 5$ it holds that $FB(s, 2^{s-1}) = 2^s-1$. 
		\item An $FB\bs(2^s-1,s, 2^{s-2} + 2^{s-4} + \left\lfloor \frac{ 2^{s/2}}{\sqrt{24}} \right\rfloor)$ code exists.
		\item For a fixed $k$ it holds that $$\lim_{s\rightarrow \infty}\frac{FB(s,k)}{s} \geq \frac{k}{\log(k+1)}.$$
		 {\item $B(s, 2^{s-1}) = 2^s-1$~\cite{Wang}.
		\item $B(s, k) = s+\Theta(\sqrt s)$ for $k = 3,4,5$~\cite{Asi,Vardy}.
		\item $B(s, k) = s+\cO(\sqrt s \log s)$ for $k >6$~\cite{Vardy}}.
	\end{enumerate}
\end{theorem}

Note that the result from Theorem~\ref{theo:0}(d) improves upon the result of $FB\bs(2^s-1,s, 2^{s-2} + 2^{s-4} + 1)$ functional batch codes which was derived from a WOM codes construction by Godlewski~\cite{Godlewski}. This is the best-known result concerning the number of queries when the number of information bits is $s$ and the number of encoded bits is $2^s-1$.

The goal of this paper is to improve some of the results summarized in Theorem~\ref{theo:0}. 
The result in $(c)$ holds for $s\leq 5$, and it was conjectured in~\cite{Zhang} that it holds for all positive values of $s$. 
\begin{conjecture}\cite{Zhang}\label{conj:1}
	For all $s > 5$,  $FB(s, 2^{s-1}) = 2^s-1$.
\end{conjecture}

The reader can notice the gap between Conjecture~\ref{conj:1} and the result in Theorem~\ref{theo:0}$(d)$. More precisely,~\cite{Zhang} assures that an $FB\bs(2^s-1,s, 2^{s-2} + 2^{s-4} +  \left\lfloor \frac{ 2^{s/2}}{\sqrt{24}} \right\rfloor)$ code exists, and the goal is to determine whether an $FB\bs(2^s-1,s, 2^{s-1})$ code exists. This paper takes one more step in establishing this conjecture. Specifically, the best-known value of the number of requested bits $k$ is improved for the case of $s$ information bits and  $2^s-1$ encoded bits. The next theorem summarizes the contributions of this paper. 
\begin{theorem}\label{theo:1} 
For a positive integer $s$, the following constructions exist:
\begin{enumerate}[(a)]
	\item A construction of $FB\bs(2^s-1,s,\lfloor\frac{2}{3}\cdot 2^{s-1} \rfloor)$ codes.
	\item A construction of  $$FB\bs(2^s +\lceil(3\alpha-2)\cdot2^{s-2}\rceil -1 ,s,\lfloor\alpha\cdot 2^{s-1}\rfloor)$$ codes where $2/3  \leq \alpha \leq 1$.  
	\item A construction of $FB\bs(2^{s+1}-2,s,2^s)$ codes.
	 {\item A construction of $FB\bs(2^s-1,s,\lfloor\frac{5}{6}\cdot 2^{s-1} \rfloor - s)$ codes.}
\end{enumerate}	
\end{theorem}

We now explain the improvements of the results of Theorem~\ref{theo:1}.  The construction in Theorem~\ref{theo:1}$(a)$ improves upon the result from Theorem~\ref{theo:0}$(d)$, where the supported number of requests increases from $\frac{1}{2}2^{s-1} + 2^{s-4} + \left\lfloor \frac{ 2^{s/2}}{\sqrt{24}} \right\rfloor$ to $\lfloor\frac{2}{3}\cdot 2^{s-1} \rfloor$. Note that by taking $\alpha =2/3$ in the result of Theorem~\ref{theo:1}$(b)$, we immediately get the result of $(a)$. However, for simplicity of the proof, we first show the construction for $(a)$ separately, and afterwards, add its extension.  { The result of Theorem~\ref{theo:1}$(d)$ is based on the result of Theorem~\ref{theo:1}$(a)$ and improves it to $\lfloor\frac{5}{6}\cdot 2^{s-1} \rfloor-s$ requests.} Moreover, according to the second result of Theorem~\ref{theo:0}$(b)$ if $t=1$ then $FP(s,2^s) \leq  2^{s+1} -2$. Based on the result in~\cite{Fazeli} it holds that $FP(s,2^s) \geq 2^{s+1} -2$. Therefore,  $FP(s,2^s) = 2^{s+1} -2$. The construction in Theorem~\ref{theo:1}$(c)$ extends this result to functional batch codes by showing that $FB(s,2^s) \leq  2^{s+1} -2$, and again, combining the result from~\cite{Fazeli}, it is deduced that $FB(s,2^s) = 2^{s+1} -2$.

A special family of matrices that will be used extensively in the paper are the generator matrices of Hadamard codes~\cite{Arora}, as defined next. 
\begin{definition}\label{HG_matrices}
	A matrix  $G = [\bfg_0,\bfg_1,\dots,\bfg_{2^s-1}]$ 
	of order $s\times 2^s$ over $\F_2$ such that
	$ \{\bfg_0,\bfg_1,\dots,\bfg_{2^s-1}\} = \F^s_2 $
	is called a \textit{Hadamard generator matrix} and in short \textit{\HG}.
\end{definition}

We will use \HGs as the generator matrices of the linear codes that will provide the constructions used in establishing Theorem~\ref{theo:1}. 
More specifically, given a linear code defined by a generator \HGG $G$ of order $s\times n$ and a request $M$ of order $s\times k$, we will show an algorithm that finds a solution for $M$. This solution will be obtained by rearranging the columns of $G$ and thereby generating a new \HGG $G'$. 
This solution is obtained by showing all the disjoint recovery sets for the request $M$, with respect to indices of columns of $G'$.
Although such a solution is obtained with respect to $G'$ instead of $G$, it can be easily adjusted to $G$ by relabeling the indices of the columns. Thus, any \HGG whose column indices are partitioned to recovery sets for $M$ provides a solution.
Note that \HGs store the all-zero column vector. Such a vector will help us to simplify the construction of the algorithm and will be removed at the end of the algorithm. 

\begin{definition}\label{def:6.1}
	Let $ M  = [\bfv_0, \bfv_1, \dots,\bfv_{n/2-1} ]$ be a request of order $s\times n/2$, where $n=2^s$. The matrix $M$ has a \textbf{Hadamard solution} if there  exists an \HGG  	$G = [\bfg_0,\bfg_1,\dots,\bfg_{n-1}]$ of order $s\times n$ such  that for all  $i\in [n/2]$,
	\begin{align*}
	&\bfv_i = \bfg_{ 2i } + \bfg_{2i+1}.
	\end{align*}
	In this case, we say that $G$ is a Hadamard solution for $M$.
\end{definition}

Next, an example is shown.
\begin{example} For $s = 3$, let
	\[G = 
	\begin{blockarray}{cccccccc}
	\bfg_0 & \bfg_1 & \bfg_2 & \bfg_3 & \bfg_4 & \bfg_5& \bfg_6& \bfg_7 \\
	\begin{block}{(cccccccc)}
	0& 1 & 0 & 1 & 0 & 1 & 0 & 1 \\
	0& 0 & 1 & 1 & 0 & 0 & 1 & 1 \\
	0& 0 & 0 & 0 & 1 & 1 & 1 & 1 \\
	\end{block}
	\end{blockarray}.
	\]	
	be an \HG. Given a request,
	\[M = 
	\begin{blockarray}{cccccccc}
	\bfv_0 & \bfv_1 & \bfv_2 & \bfv_3   \\
	\begin{block}{(cccccccc)}
	\textcolor{blue}{0} & \textcolor{orange}{0} & \textcolor{red}{0} & \textcolor{brown}{0} \\
	\textcolor{blue}{0}& \textcolor{orange}{0} & \textcolor{red}{0} & \textcolor{brown}{0}   \\
	\textcolor{blue}{1}& \textcolor{orange}{1} & \textcolor{red}{1} & \textcolor{brown}{1}   \\
	\end{block}
	\end{blockarray}
	\]
	a Hadamard solution for this request may be
	\[G' = 
	\begin{blockarray}{cccccccc}
	\bfg'_0 & \bfg'_1 & \bfg'_2 & \bfg'_3 & \bfg'_4 & \bfg'_5& \bfg'_6& \bfg'_7 \\
	\begin{block}{(cccccccc)}
	 \textcolor{blue}{0} &\textcolor{blue}{0} & \textcolor{orange}{0} & \textcolor{orange}{0} & \textcolor{red}{1} & \textcolor{red}{1}&  \textcolor{brown}{1} &  \textcolor{brown}{1}\\
	 \textcolor{blue}{0} &\textcolor{blue}{0}& \textcolor{orange}{1} & \textcolor{orange}{1} & \textcolor{red}{0} & \textcolor{red}{0} &  \textcolor{brown}{1} & \textcolor{brown}{1}  \\
	  \textcolor{blue}{0} &\textcolor{blue}{1}& \textcolor{orange}{0} & \textcolor{orange}{1} & \textcolor{red}{0} & \textcolor{red}{1} &  \textcolor{brown}{0} & \textcolor{brown}{1} \\
	\end{block}
	\end{blockarray}.
	\]
\end{example}

Lastly, for the convenience of the reader, the relevant notations and terminology that will be used throughout the paper is summarized
in Table~\ref{tab:cql}.

\begin{table}[!h]
	\caption{Table of Definitions and Notations}\label{tab:cql}
	\begin{center}	 \vspace{-4ex}
	\begin{tabular}{ccc}
			\hline
			Notation & Meaning & Remarks   \\
			\hline
			$FB\bs(n,s,k)$   &  \hspace{-3ex}A func. $k$-batch code of length $n$ and dimension $s$ \hspace{-3ex} &  Sec.~\ref{sec:defs}  \\
			$B\bs(n,s,k)$   &  \hspace{-3ex}A $k$-batch code of length $n$ and dimension $s$ \hspace{-3ex} &  Sec.~\ref{sec:defs}  \\
			$R_i$ &  \hspace{-3ex}The $i$-th recovery set \hspace{-3ex}& Sec.~\ref{sec:defs}  \\
			$(\cG,\cB,\cR)$ &  \hspace{-3ex}A triple-set  \hspace{-3ex}& Def.~\ref{def:2}  \\
			$\cM(\cG,\cB,\cR)$ &  \hspace{-3ex}A triple-matrix of $M$  \hspace{-3ex}& Def.~\ref{def:2}  \\
			$\bfe$ & \hspace{-3ex} A unit vector of length $s$ with $1$ at its last index \hspace{-3ex} & Sec.~\ref{sec:cons1}  \\
			$M$ &\hspace{-3ex}  A request matrix \hspace{-3ex} & Sec.~\ref{sec:cons1}  \\
			$\bfv_i,\bfw_i$ & \hspace{-3ex} The $i$-th request/column vector in $M,\cM$ \hspace{-3ex} & Sec.~\ref{sec:cons1}  \\	 
			$G$ & \hspace{-3ex} An \HG \hspace{-3ex} & Sec.~\ref{sec:cons1} \\
			$\bfg_i$ &  \hspace{-3ex}A column vector in $G$ representing the $i$-th server \hspace{-3ex} & Sec.~\ref{sec:cons1}  \\ 		
			$\bfG_\bfx(G)$ & \hspace{-3ex} An $\bfx$-type graph of $G$ \hspace{-3ex} & Def.~\ref{def:5}  \\
 			$\bfC_\bfx(G)$ &\hspace{-3ex}  The partition of simple cycles of $\bfG_\bfx(G)$ \hspace{-3ex} & Def.~\ref{def:5}  \\
			$P_{\bfx}(\bfg_i,\bfg_j)$ &\hspace{-3ex}   A simple path between $\bfg_i$ and $\bfg_j$ in $\bfG_\bfx(G)$ \hspace{-3ex} & Def.~\ref{def:7}  \\
			$d_{P_{\bfx}}(\bfg_{t},\bfg_{m})$ & \hspace{-3ex}  The sub-length from $\bfg_t$ to $\bfg_m$ in $P_{\bfx}(\bfg_i,\bfg_j)$ \hspace{-3ex} & Def.~\ref{def:7}  \\			
			$\cF_{\bfx}(\bfg_i,\bfg_j)$ &\hspace{-3ex}  A reordering function for a good-path $P_{\bfx}(\bfg_i,\bfg_j)$ \hspace{-3ex} & Def.~\ref{def:7}  \\			 
			\hline
		\end{tabular}
	\end{center}
\end{table}

\section{A Construction of $FB\bs(2^s-1,s,\lfloor\frac{2}{3}\cdot 2^{s-1} \rfloor)$ Codes}\label{sec:cons1}
	In this section a construction of $FB\bs(2^s-1,s,\lfloor\frac{2}{3}\cdot 2^{s-1} \rfloor)$ codes is presented. 	Let the request $M$ be denoted by $$M = [\bfv_0, \bfv_1, \dots,\bfv_{\lfloor\frac{2}{3}\cdot 2^{s-1} \rfloor-1}].$$
	Let $\bfe = (0,0,\dots,0,1) \in \F^s_2$ be the unit vector with 1 at its last index. The solution for the request $M$ will be derived by using two algorithms as will be presented in this section. We start with several definitions and tools that will be used in these algorithms. 
	 \begin{definition}\label{def:2}
 	Three sets $\cG,\cB,\cR\subseteq [2^{s-1}]$ are called a \textbf{triple-set} (the good, the bad, and the redundant), and are denoted by $(\cG,\cB,\cR)$, if the following properties hold,
 	\begin{align*}
 	&\cG\subseteq \Big[\Big\lfloor\frac{2}{3}\cdot 2^{s-1} \Big\rfloor\Big],   \\
 	&\cB = \Big[\Big\lfloor\frac{2}{3}\cdot 2^{s-1} \Big\rfloor\Big] \setminus \cG,\\
 	&\cR = [2^{s-1}]\setminus \Big(\cG\cup \cB\cup\{2^{s-1}-1\}\Big).
 	\end{align*}
 	Given a matrix
 	$M = [\bfv_0, \bfv_1, \dots,\bfv_{\lfloor\frac{2}{3}\cdot 2^{s-1} \rfloor-1}]$
 	of order $s\times \lfloor\frac{2}{3}\cdot~2^{s-1} \rfloor$, the matrix $\cM(\cG,\cB,\cR) = [\bfw_0, \bfw_1, \dots,\bfw_{2^{s-1}-1}]$ of order  $s\times~2^{s-1}$ is referred as a \textbf{triple-matrix of} $M$ if it holds that
 	\begin{align*}
 	\bfw_t =
 	\begin{cases}
 	\bfv_t & t\in \cG\\
 	\bfv_t + \bfe  & t\in \cB \\
 	\bfe  & t\in \cR
 	\end{cases}.
 	\end{align*}
 	
 \end{definition}
	Note that, we did not demand anything about the vector $\bfw_{2^{s-1}-1}$, i.e., it can be any binary vector of length $s$.
	Furthermore, by Definition~\ref{def:2}, the set $\cB$ uniquely defines the triple-set $(\cG,\cB,\cR)$. We proceed with the following claim.
 	\begin{claim}\label{claim:1}
 		For any triple-set $(\cG,\cB,\cR)$ if $|\cB| \leq  \lfloor\frac{1}{3}\cdot 2^{s-1}  \rfloor$ then $|\cB| \leq |\cR|$.
 	\end{claim}
 
 	\begin{IEEEproof}
	According to the definition of $(\cG,\cB,\cR)$ and since $\cG\cup \cB = [\lfloor\frac{2}{3}\cdot 2^{s-1} \rfloor]$ it holds that 
	\begin{align*}
	|\cR| &= \Big|[2^{s-1}]\setminus \Big(\cG\cup \cB\cup\{2^{s-1}-1\}\Big)\Big|\\
	&= 2^{s-1} -  \Big\lfloor\frac{2}{3}\cdot 2^{s-1} \Big\rfloor -1.
	\end{align*}
	Thus, in order to prove that $|\cB| \leq |\cR|$, since  $|\cB| \leq  \lfloor\frac{1}{3}\cdot 2^{s-1}  \rfloor$, we will prove inequality $(a)$ in 
	\begin{align*}
	|\cR| = 2^{s-1} -  \Big\lfloor\frac{2}{3}\cdot 2^{s-1} \Big\rfloor -1 \stackrel{(a)}{\geq} \Big\lfloor\frac{1}{3}\cdot 2^{s-1} \Big\rfloor \geq |\cB|.
	\end{align*}
	This inequality equivalent to
	\begin{align*}
	2^{s-1}  -1 \geq   \Big\lfloor\frac{2}{3}\cdot 2^{s-1} \Big\rfloor  + \Big\lfloor\frac{1}{3}\cdot 2^{s-1} \Big\rfloor.
	\end{align*}
	We separate the proof for the following two cases.\\
	\textbf{Case 1:} If $s$ is even, then 
	$$2^s \equiv 1 (\bmod 3),~~~~2^{s-1} \equiv 2 (\bmod 3).$$ Thus,
	\begin{align*}
	   &\Big\lfloor\frac{2}{3}\cdot 2^{s-1} \Big\rfloor  + \Big\lfloor\frac{1}{3}\cdot 2^{s-1} \Big\rfloor =
	   \Big\lfloor\frac{2^s}{3}\Big\rfloor  + \Big\lfloor\frac{2^{s-1}}{3} \Big\rfloor\\
	   &=\frac{ 2^{s}-1}{3}  +  \frac{ 2^{s-1}-2}{3} = \frac{3\cdot 2^{s-1} - 3}{3} = 2^{s-1}  -1.
	\end{align*}\\
	\textbf{Case 2:} If $s$ is odd, then	
	$$2^s \equiv 2 (\bmod 3),~~~~2^{s-1} \equiv 1 (\bmod 3).$$ Thus,
	\begin{align*}
	&\Big\lfloor\frac{2}{3}\cdot 2^{s-1} \Big\rfloor  + \Big\lfloor\frac{1}{3}\cdot 2^{s-1} \Big\rfloor =
	\Big\lfloor\frac{2^s}{3}\Big\rfloor  + \Big\lfloor\frac{2^{s-1}}{3} \Big\rfloor\\
	&=\frac{ 2^{s}-2}{3}  +  \frac{ 2^{s-1}-1}{3} = \frac{3\cdot 2^{s-1} - 3}{3} = 2^{s-1}  -1.
	\end{align*}\\
	Therefore, it is deduced that in both cases if $|\cB| \leq  \lfloor\frac{1}{3}\cdot 2^{s-1}  \rfloor$ then $|\cB|\leq |\cR|$.
	\end{IEEEproof}
	As mentioned above, our strategy is to construct two algorithms. We start by describing the first one which is the main algorithm. This algorithm receives as an input the request $M$ and outputs a set $\cB$ and a Hadamard-solution for some triple-matrix $\cM(\cG,\cB,\cR)$ of $M$. Using the matrix $\cM(\cG,\cB,\cR)$, it will be shown how to derive the solution for $M$. This connection is established in the next lemma. For the rest of this section we denote $n=2^s$ and for our ease of notations both of them will be used.
	
	\begin{lemma}\label{lemma:5}
		If there is a Hadamard solution for  $\cM(\cG,\cB,\cR)$ such that  $|\cB| \leq  \lfloor\frac{1}{3}\cdot 2^{s-1}  \rfloor$, then there is a solution for 	$M = [\bfv_0, \bfv_1, \dots,\bfv_{\lfloor\frac{2}{3}\cdot 2^{s-1} \rfloor-1}]$.
	\end{lemma}
	
	\begin{IEEEproof}
	Let the \HGG $G = [\bfg_0,\bfg_1,\dots,\bfg_{n-1}]$ be a  Hadamard solution for 
	$\cM(\cG,\cB,\cR)$. Our goal is to form all disjoint recovery sets $R_t$ for $t\in \cG\cup\cB = [ \lfloor\frac{2}{3}\cdot 2^{s-1}  \rfloor ]$ for $M$. Since $G$ is a Hadamard solution for $\cM(\cG,\cB,\cR)$,	for all $t\in [2^{s-1}]$, it holds that
	\begin{align*}
		 \bfw_t= \bfg_{2t} + \bfg_{2t+1}.
	\end{align*}
	By definition of $\cM(\cG,\cB,\cR)$
	 \begin{align*}
	\bfw_t =
	\begin{cases}
	\bfv_t & t\in \cG\\
	\bfv_t + \bfe  & t\in \cB \\
	\bfe  & t\in \cR
	\end{cases}.
	\end{align*}
	Thus,  if $ t\in \cG$ then
	\begin{align*}
	&\bfv_t = \bfw_t= \bfg_{2t} + \bfg_{2t+1},
	\end{align*}
	 and each recovery set for $\bfv_t$ is of the form $R_t = \{2t,2t+1\}$.
	 If $t\in \cB$ then
	\begin{align*}
	&\bfv_t +\bfe = \bfw_t= \bfg_{2t} + \bfg_{2t+1},
	\end{align*}
	and if $t'\in \cR$ then
	\begin{align*}
	&\bfe = \bfw_{t'}= \bfg_{2t'} + \bfg_{2t'+1}.
	\end{align*}
	Therefore, for all $t\in\cB$ and $t'\in \cR$,
	\begin{align*}
	\bfv_t = \bfg_{2t} + \bfg_{2t+1} + \bfg_{2t'} + \bfg_{2t'+1}.
	\end{align*}
	By Claim~\ref{claim:1}, since $|\cB| \leq  \lfloor\frac{1}{3}\cdot 2^{s-1}  \rfloor$, it holds that $|\cB| \leq~|\cR| $. Thus, for all $t\in \cB$, each recovery set $R_t$ for $\bfv_t$ will have a different $t' \in \cR$ such that $$ R_t = \{2t,2t+1,2t',2t'+1\}.$$  
	\end{IEEEproof}
	In Lemma~\ref{lemma:5}, it was shown that obtaining $\cM(\cG,\cB,\cR)$ which holds $|\cB| \leq  \lfloor\frac{1}{3}\cdot 2^{s-1}  \rfloor$ provides a solution for $M$. Therefore, if the first algorithm outputs a set $\cB$ for which $|\cB| \leq  \lfloor\frac{1}{3}\cdot 2^{s-1}  \rfloor$, then the solution for $M$ is easily derived. Otherwise, the first algorithm outputs a set $\cB$ such that $|\cB| > \lfloor\frac{1}{3}\cdot 2^{s-1}  \rfloor$. In this case, the second algorithm will be used in order to reduce the size of the set $\cB$ to be at most $\lfloor\frac{1}{3}\cdot 2^{s-1}  \rfloor$. For that, more definitions are required, and will be presented in the next section.

\subsection{Graph Definitions}

In the two algorithms of the construction, we will use undirected graphs, simple paths, and simple cycles that will be defined next. These graphs will be useful to represent the \HGG $G$ in some graph representation and to make some swap operations on its columns.

\begin{definition}\label{def:9}
	An \textbf{undirected graph} or simply a \textbf{graph} will be denoted by $\bfG=(V,E)$,  where $V=\{u_0,u_1,\ldots,u_{m-1}\}$ is its set of $m$ nodes (vertices) and  $E\subseteq\{\{u_i,u_j\}~|~u_i,u_j \in V\}$ is its edge set. A finite \textbf{simple path} of length $\ell$ is a sequence of distinct edges $e_0,e_1,\dots,e_{\ell-1}$ for which there is a sequence of vertices $u_{i_0},u_{i_1},\dots,u_{i_{\ell}}$ such that $e_j = \{u_{i_j},u_{i_{j+1}}\}, j \in [\ell]$. A \textbf{simple cycle} is a simple path in which $u_{i_0} = u_{i_\ell}$. The \textbf{degree} of a node $u_i$ is the number of edges that are incident to the node, and will be denoted by $\deg(u_i)$. 
\end{definition}

Note that in Definition~\ref{def:9} we did not allow parallel edges, i.e., different edges which connect between the same two nodes. 
By a slight abuse of notation, we will use graphs in which at most $2$ parallel edges are allowed between any two nodes. That implies that cycles of length $2$ may appear in the graph. In this case, we will use some notations for distinguishing between two parallel edges as will be done in the following definition. 

\begin{definition}\label{def:5}
	Given an  \HGG $G = [\bfg_0,\bfg_1,\dots,\bfg_{n-1}],$
	and a vector $\bfx \in  \F^s_2$,
	denote the \textbf{$\bfx$-type graph} $\bfG_{\bfx}(G) = (V,E_\bfx(G))$ of $G$ and $\bfx$ such that $V = \F^s_2$ and a multi-set
	\begin{align*}
	& E_\bfx(G) =   \Big\{ \{\bfg_{i},\bfg_{i} + \bfx\} ~|~ i\in [n] \Big\}\cup\Big\{ \{\bfg_{2t-1},\bfg_{2t}\} ~|~ t\in [n/2] \Big\} .
	\end{align*}
	For all $t \in [n/2]$, we say that $\bfg_{2t-1}$ and $\bfg_{2t}$ are a \textbf{pair}. An edge $\{\bfg_{2t-1},\bfg_{2t}\}$ will be called a \textbf{pair-type edge} and will be denoted by $\{\bfg_{2t-1},\bfg_{2t}\}_\mathbf{p}$. An edge $\{\bfg_{i},\bfg_{i} + \bfx\}$ will be called an \textbf{\textbf{$\bfx$}-type edge} and will be denoted by  $\{\bfg_{i},\bfg_{i} + \bfx\}_{\bfx}$.  
	Note that for any $ \bfg \in V$, it holds that $\deg(\bfg) = 2$.  Thus, the graph $\bfG_{\bfx}(G)$ has a partition of $\ell\geq 1$ disjoint simple cycles, that will be denoted by $\bfC_{\bfx}(G) = \{C_i\}^{\ell-1}_{i=0}$,  { where every $C_i$ is denoted by its set of edges.}
\end{definition}

	Note that $E_\bfx(G)$ is a multi-set since in case that $\bfg_{2t-1} = \bfg_{2t} + \bfx$, we have two parallel edges $\{\bfg_{2t-1},\bfg_{2t}\}_\mathbf{p}$ and $\{\bfg_{2t-1},\bfg_{2t}\}_\bfx$ between $\bfg_{2t-1}$ and $\bfg_{2t}$. For the following definitions assume that $G =  [\bfg_0,\bfg_1,\dots,\bfg_{n-1}]$ is an \HGG of order $s\times n$.

\begin{definition}\label{def:6}
	Given an  $\bfx$-type graph $\bfG_\bfx(G)$ such that $\bfx\in \F^s_2$, let $\bfg_i,\bfg_j$ be two vertices connected by a simple path  $P_{\bfx}(\bfg_i,\bfg_j,G)$ of length $\ell-1$  in $\bfG_\bfx(G)$ which is denoted by
	\begin{align*}
	\bfg_i = \bfg_{s_0}-\bfg_{s_1}- \dots -\bfg_{s_{\ell-1}} = \bfg_j.
	\end{align*}
	The path $P_{\bfx}(\bfg_i,\bfg_j,G)$ will be called a \textbf{good-path} if the edges $\{\bfg_{s_0},\bfg_{s_1}\}$ and $\{\bfg_{s_{\ell-2}},\bfg_{s_{\ell-1}}\}$ are both $\bfx$-type edges. 
	For all $\bfg_{t}$ and $\bfg_{m}$ on $P_{\bfx}(\bfg_i,\bfg_j,G)$, denote by $d_{P_{\bfx}}(\bfg_{t},\bfg_{m},G)$ the length of the simple sub-path from  $\bfg_{t}$ to $\bfg_{m}$ on $P_{\bfx}(\bfg_i,\bfg_j,G)$. This length will be called the \textbf{sub-length from $\bfg_t$ to $\bfg_m$ in $P_{\bfx}(\bfg_i,\bfg_j,G)$}. When the graph $G$ will be clear from the context we will use the notation $P_{\bfx}(\bfg_i,\bfg_j)$, $d_{P_{\bfx}}(\bfg_{t},\bfg_{m})$ instead of $P_{\bfx}(\bfg_i,\bfg_j,G)$, $d_{P_{\bfx}}(\bfg_{t},\bfg_{m},G)$, respectively.
\end{definition}

We next state the following claim.

\begin{claim}\label{claim:3}
	Given a good-path $P_{\bfx}(\bfg_i,\bfg_j)$ of length $\ell-1$ in $\bfG_\bfx(G)$
		\begin{align*}
	\bfg_i = \bfg_{s_0}-\bfg_{s_1}- \dots -\bfg_{s_{\ell-1}} = \bfg_j,
	\end{align*}
	 where  $\bfx \in \F^s_2$, the following properties hold. 
	\begin{enumerate}
		\item The value of $\ell$ is even.\label{claim:3_3}
		\item For all $m\in[\ell/2-1]$ the edge $\{\bfg_{s_{2m+1}},\bfg_{s_{2m+2}}\}_{_\mathbf{p}}$ is a pair-type edge.\label{claim:3_31}
		\item For all $t\in [\ell/2]$, $\bfg_{s_{2t}} = \bfg_{s_{2t+1}} + \bfx$.\label{claim:3_4}
		\item If $\bfg_i,\bfg_j$ is not a pair, then
		the pair of $\bfg_{i}$ and the pair of  $\bfg_{j}$ are not in $P_{\bfx}(\bfg_i,\bfg_j)$.\label{claim:3_1}	
	\end{enumerate}
\end{claim}

\begin{IEEEproof}
	We prove this claim as follows.
	\begin{enumerate}
		\item Since $P_{\bfx}(\bfg_i,\bfg_j)$ is a good-path, by definition the edge $\{\bfg_{s_0},\bfg_{s_1}\}_{\bfx}$ is an $\bfx$-type edge. We also know that for all $t\in [\ell]$ it holds that $\deg(\bfg_{s_t}) = 2$. Thus, the edge $\{\bfg_{s_1},\bfg_{s_2}\}_{_\mathbf{p}}$ is a pair-type edge, the edge $\{\bfg_{s_2},\bfg_{s_3}\}_{\bfx}$ is an $\bfx$-type edge, and so on. More formally, for all $t\in [\ell/2]$ the edge $\{\bfg_{s_{2t}},\bfg_{s_{2t+1}}\}_{\bfx}$ is an $\bfx$-type edge and for all $m\in[\ell/2-1]$ the edge $\{\bfg_{s_{2m+1}},\bfg_{s_{2m+2}}\}_{_\mathbf{p}}$ is a pair-type edge. Since the last edge $\{\bfg_{s_{\ell-2}},\bfg_{s_{\ell-1}}\}_\bfx$ is also an $\bfx$-type edge, we deduce that $\ell-1$ is odd or equivalently $\ell$ is even.
		\item The proof of this part holds due to a).
		\item In a) we proved that for all $t\in [\ell/2]$ the edge $\{\bfg_{s_{2t}},\bfg_{s_{2t+1}}\}_{\bfx}$ is an $\bfx$-type edge. Thus, by definition $\bfg_{s_{2t}} = \bfg_{s_{2t+1}} + \bfx.$
		\item Let $\bfg_m$ be a pair of $\bfg_i$ and we will prove that $\bfg_m \notin P_{\bfx}(\bfg_i,\bfg_j)$. Note that $\bfg_m\neq \bfg_j$ and $\deg(\bfg_m) = 2$. Therefore, if  $\bfg_m \in P_{\bfx}(\bfg_i,\bfg_j)$, then $\bfg_i$ has to appear more than once in $ P_{\bfx}(\bfg_i,\bfg_j)$. This is in contradiction to the fact that  $ P_{\bfx}(\bfg_i,\bfg_j)$ is a simple path.
	\end{enumerate}
\end{IEEEproof}

Another useful property on good-paths in $\bfx$-type graphs is proved in the next claim.
\begin{claim}\label{claim:3_2}
	If $\bfg_i,\bfg_j$ is a pair, then there is a good-path $P_{\bfx}(\bfg_{i},\bfg_{j})$ in $\bfG_{\bfx}(G)$.
\end{claim}

\begin{IEEEproof}
	 We know that all nodes in $\bfG_{\bfx}(G)$ are of degree $2$. Therefore, there is a simple cycle in $C \in \bfC_{\bfx}(G)$ including the edges $\{\bfg_{i},\bfg_m\}_\bfx$ and $\{\bfg_{j},\bfg_p\}_\bfx$ for some $m,p \in [n]$, and the edge $\{\bfg_{i},\bfg_{j}\}_{_\mathbf{p}}$. By removing the edge $\{\bfg_{i},\bfg_{j}\}_{_\mathbf{p}}$ from  $C$ we get a simple path $P$ starting with the edge $\{\bfg_{i},\bfg_m\}_\bfx$ and ending with the edge $\{\bfg_{j},\bfg_p\}_\bfx$. Thus, by definition, $P$ is a  good-path $P_{\bfx}(\bfg_{i},\bfg_{j})$.
\end{IEEEproof}

The next definition will be used for changing the order of the columns in $G$.
\begin{definition}\label{def:7}
	Let $\cH_{s}$ be the set of all \HGs of order $s\times n$. Let $\cP_{s} \subseteq \F^{s}_2\times \F^{s}_2$ be the set of all couples of column vectors $\bfg_m,\bfg_p$ of $G$ such that there is a good-path  $P_{\bfx}(\bfg_m,\bfg_p)$.
	For every two column vectors $\bfg_i,\bfg_j$ with a good-path $P_{\bfx}(\bfg_i,\bfg_j)$ of length $\ell-1$ in $\bfG_\bfx(G)$
			\begin{align*}
	\bfg_i = \bfg_{s_0}-\bfg_{s_1}- \dots -\bfg_{s_{\ell-1}} = \bfg_j,
	\end{align*}
	 denote the \textbf{reordering function} $\cF_\bfx: \cP_{s} \times  \cH_s \rightarrow \cH_s $ that generates an \HGG $\cF_{\bfx} (\bfg_i,\bfg_j,G)$ from $G$ by adding $\bfx$ to every column $\bfg_{s_m},m\in [\ell]$. We will use the notation $\cF_{\bfx} (\bfg_i,\bfg_j)$ for shorthand.
\end{definition}

The following claim proves that the function $\cF_\bfx$ is well defined.

\begin{claim}
	The matrix $\cF_{\bfx} (\bfg_i,\bfg_j)$ is an \HGG of order $s\times n$.
\end{claim}

\begin{IEEEproof}
	Let $ P_{\bfx}(\bfg_i,\bfg_j)$ be a good-path of length $\ell-1$ in $\bfG_\bfx(G)$ denoted by	
	\begin{align*}
	\bfg_i = \bfg_{s_0}-\bfg_{s_1}- \dots -\bfg_{s_{\ell-1}} = \bfg_j.
	\end{align*}
	By using the function $\cF_{\bfx} (\bfg_i,\bfg_j) $, the vector  $\bfx$ is added to every column $\bfg_{s_m},m\in [\ell]$.
	In Claim~\ref{claim:3}\eqref{claim:3_4} it was shown that for all $t \in [\ell/2]$,
	$$\bfg_{s_{2t}} = \bfg_{s_{2t+1}} + \bfx.$$
	Therefore, adding $\bfx$ to all the columns $\bfg_{s_m}, m\in [\ell]$, is equivalent to swapping the column vectors  $\bfg_{s_{2t}}, \bfg_{s_{2t+1}}$ for all $t\in [\ell/2]$ in $G$. Since after rearranging the columns of $G$, it is still an \HG, it is deduced that $\cF_{\bfx} (\bfg_i,\bfg_j)$ is an \HG.
\end{IEEEproof}

To better explain these definitions and properties, the following example is presented.
\begin{example} For $s = 3$, let $G$ be the following \HG 
	\[G = 
	\begin{blockarray}{cccccccc}
	\bfg_0 & \bfg_1 & \bfg_2 & \bfg_3 & \bfg_4 & \bfg_5& \bfg_6& \bfg_7 \\
	\begin{block}{(cccccccc)}
	\textcolor{blue}{0} & \textcolor{blue}{1} & \textcolor{orange}{0} & \textcolor{orange}{1} & \textcolor{red}{0} & \textcolor{red}{1}&  \textcolor{brown}{0} &  \textcolor{brown}{1}  \\
	\textcolor{blue}{0} & \textcolor{blue}{0}& \textcolor{orange}{1} & \textcolor{orange}{1} & \textcolor{red}{0} & \textcolor{red}{0} &  \textcolor{brown}{1} & \textcolor{brown}{1}  \\
	\textcolor{blue}{0} & \textcolor{blue}{0}& \textcolor{orange}{0} & \textcolor{orange}{0} & \textcolor{red}{1} & \textcolor{red}{1} &  \textcolor{brown}{1} & \textcolor{brown}{1} \\
	\end{block}
	\end{blockarray}.
	\]
	Let $\bfx = (1,0,1)$. The matrix $\bfG_\bfx(G)$ will be defined as in Figure~\ref{fig:1}.
	\begin{figure}[h!] 
		\centering
		\includegraphics[width=50mm]{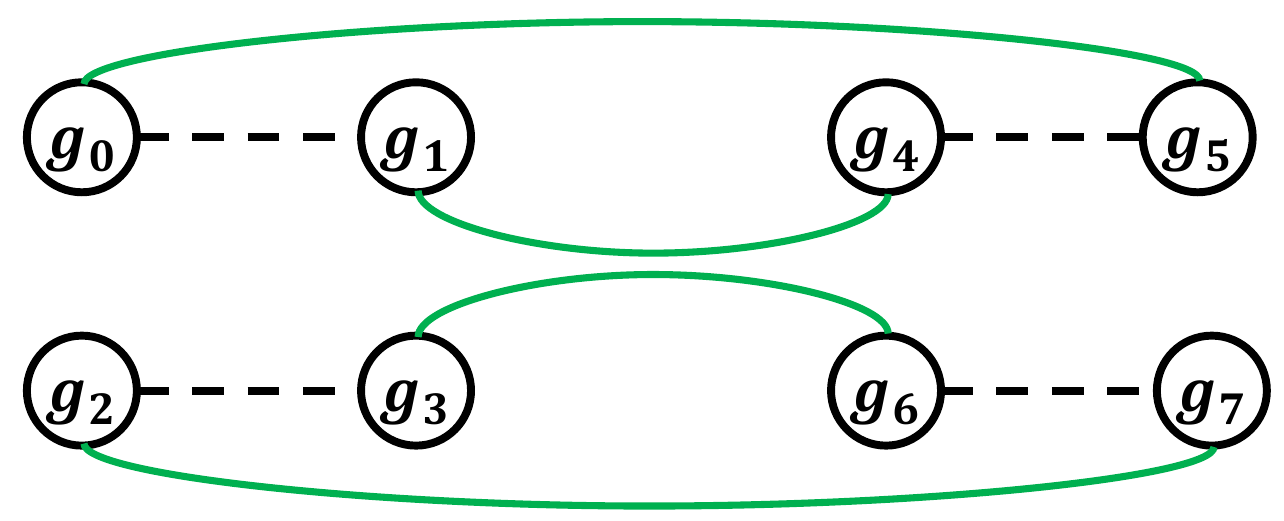} 
		\caption{The $\bfG_\bfx(G)$ graph. The green edges are the $\bfx$-type edges and the dashed edges are the pair-type edges.}\label{fig:1}
	\end{figure}
	Note that in this case, the graph $\bfG_\bfx(G)$ is partitioned into two disjoint cycles.
	While the path $ \bfg_0 - \bfg_5 $
	is a good-path between $\bfg_0$ and $\bfg_5$, the path 
	$$ \bfg_0 - \bfg_1 - \bfg_4 - \bfg_5 $$ 
	is not a good-path between $\bfg_0$ and $\bfg_5$. Note that there is no good-path between $\bfg_0$ and $\bfg_4$. Let $P_{\bfx}(\bfg_0,\bfg_1)$ be the good-path between $\bfg_0$ and $\bfg_1$,
	$$ \bfg_0 - \bfg_5 - \bfg_4 - \bfg_1.$$ 
	Thus, $G' =\cF_{\bfx}(\bfg_0,\bfg_1)$ is the following \HG
	\[G'= 
	\begin{blockarray}{cccccccc}
	\bfg'_0 & \bfg'_1 & \bfg'_2 & \bfg'_3 & \bfg'_4 & \bfg'_5& \bfg'_6 & \bfg'_7 \\
	\begin{block}{(cccccccc)}
	\textcolor{blue}{1} & \textcolor{blue}{0} & \textcolor{orange}{0} & \textcolor{orange}{1} & \textcolor{red}{1} & \textcolor{red}{0}&  \textcolor{brown}{0} &  \textcolor{brown}{1} \\
	\textcolor{blue}{0} & \textcolor{blue}{0}& \textcolor{orange}{1} & \textcolor{orange}{1} & \textcolor{red}{0} & \textcolor{red}{0} &  \textcolor{brown}{1} & \textcolor{brown}{1} \\
	\textcolor{blue}{1} & \textcolor{blue}{1}& \textcolor{orange}{0} & \textcolor{orange}{0} & \textcolor{red}{0} & \textcolor{red}{0} &  \textcolor{brown}{1} & \textcolor{brown}{1} \\
	\end{block}
		\bfg_5 & \bfg_4 & \bfg_2 & \bfg_3 & \bfg_1 & \bfg_0& \bfg_6 & \bfg_7 \\
	\end{blockarray},
	\]
	with a new graph $\bfG_\bfx(G')$ as depicted in Figure~\ref{fig:2}.
	\begin{figure}[h!] 
	 
	\subfigure[The graph $\bfG_\bfx(G')$ represented by the nodes $\{\bfg_0,\bfg_1,\ldots,\bfg_7\}$.]{\includegraphics[width=43.2mm]{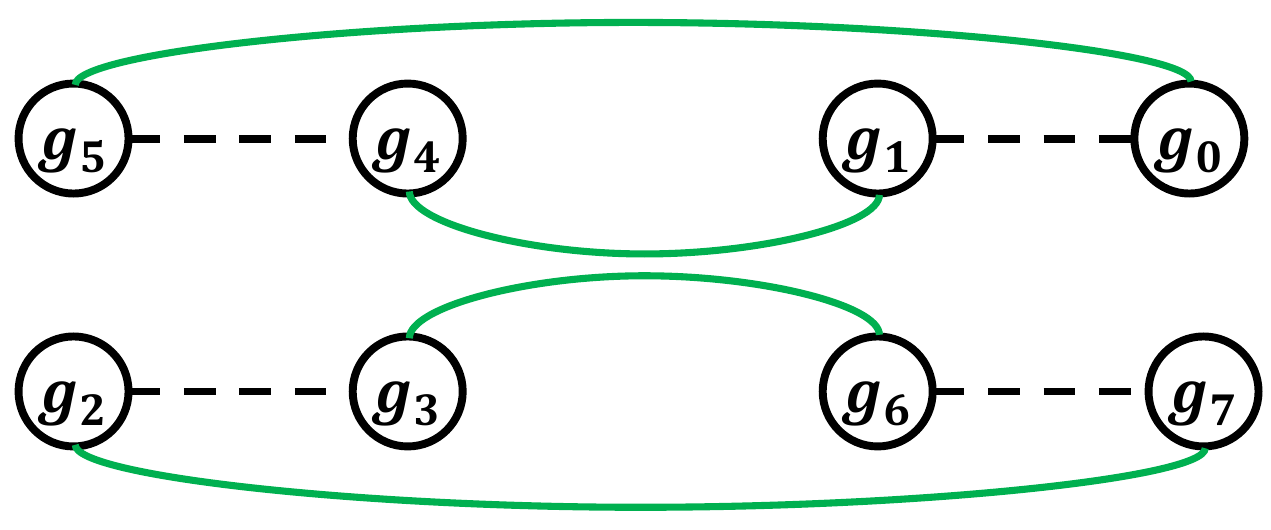}} 
 \subfigure[The graph $\bfG_\bfx(G')$ represented by the nodes $\{\bfg_0',\bfg_1',\ldots,\bfg_7'\}.$]{\includegraphics[width=43.2mm]{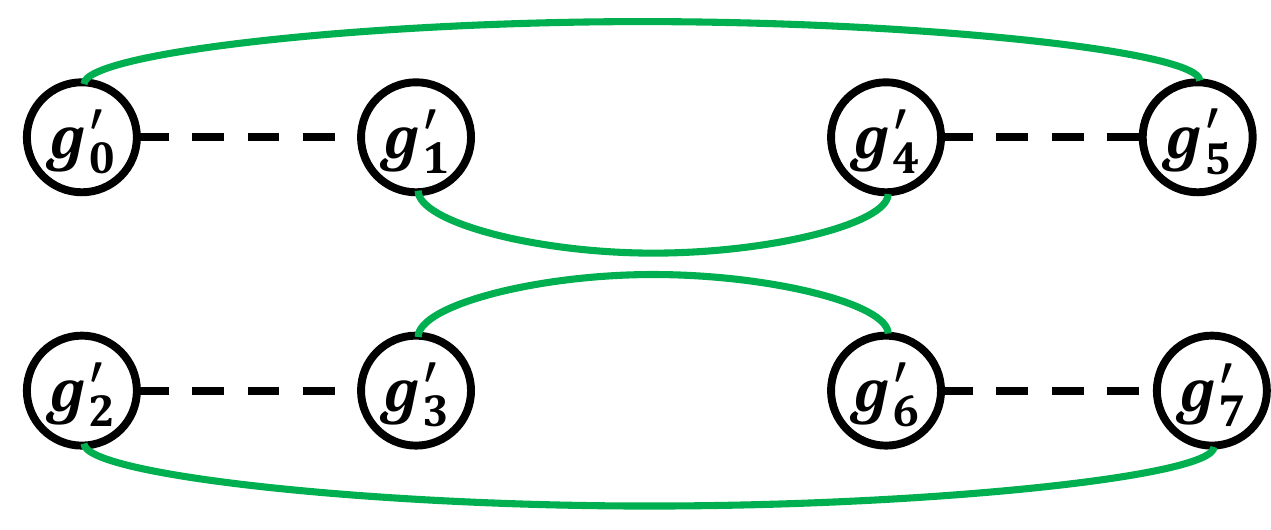}} 
	\centering
	\caption{The graph $\bfG_\bfx(G')$.}\label{fig:2}
\end{figure}
\end{example}

The next lemma shows a very important property that will be used in the construction of the first algorithm. This algorithm will have a routine of $\lfloor\frac{2}{3}\cdot 2^{s-1} \rfloor$ iterations. In iteration $t\leq \lfloor\frac{2}{3}\cdot 2^{s-1} \rfloor$, we will modify the order of the column vectors of $G$ such that only the sums $\bfg_{2t}+\bfg_{2t+1}$ and $\bfg_{n-2}+\bfg_{n-1}$ will be changed by $\bfx \in \F^s_2$, and all other sums $\bfg_{2p}+\bfg_{2p+1}$ where $p\neq t,n/2-1$ will remain the same. 
The goal on the $t$-th iteration is to get that
$$\bfg_{2t}+\bfg_{2t+1} =	\bfv_t + \mathds{1}_t \bfe,$$ 
where $\mathds{1}_t \in \{0,1\}$ and remember that $\bfe = (0,0,\dots,0,1) \in \F^s_2$. 

\begin{lemma}\label{lemma:8}
	Let $P_{\bfx}(\bfg_{r_1},\bfg_{r_2})$ be a good-path  in $\bfG_\bfx(G)$ where $\bfx \in \F^s_2$ and $r_1,r_2\in [n]$ such that
	$\bfg_{r_1},\bfg_{r_2}$ is not a pair. If
	 $r_1\in \{2i , 2i+1 \}$, $r_2\in \{2j , 2j+1 \}$ (and note that $i\neq j$),
	then, the \HG
	$$G' = \cF_{\bfx}(\bfg_{r_1},\bfg_{r_2})  = [\bfg'_0,\bfg'_1,\dots,\bfg'_{n-1}]$$
	satisfies the following equalities
	\begin{align*}
	&\bfg'_{2p} + \bfg'_{2p+1} = \bfg_{2p} + \bfg_{2p+1} + \bfx &  p\in \{i,j\},\\
	&\bfg'_{2p} + \bfg'_{2p+1} = \bfg_{2p} + \bfg_{2p+1} &  p\neq i,j,
	\end{align*}
	where $p\in [n/2]$.
\end{lemma}
 \begin{IEEEproof}
	We prove this lemma only for $r_1= 2i $ and $r_2=2j$ where $i<j$ while all other cases are proved similarly.  
	Suppose that the good-path $P_{\bfx}(\bfg_{ 2i },\bfg_{2j})$ is of length $\ell-1$ and denote it by
	$$ \bfg_{ 2i } = \bfg_{s_0}-\bfg_{s_1}- \dots -\bfg_{s_{\ell-1}} = \bfg_{2j}.$$
	Let $S$ be the set $S = \{s_0,s_1,\dots,s_{\ell-1}\}$.
	Let 
	$$G' = [\bfg'_0,\bfg'_1,\dots,\bfg'_{n-1}]$$ 
	be an \HGG of order $s\times 2^{s}$  generated by applying $\cF_{\bfx}(\bfg_{ 2i },\bfg_{2j},G)$. 
	Thus, it is deduced that for all $m\in [n]$ 
	\begin{align*}
	&\bfg'_m = \bfg_m  & \textrm{if~} m\notin S, \\
	&\bfg'_m = \bfg_m + \bfx  & ~~\textrm{ if~} m\in S.	 
	\end{align*}	
	Since $P_{\bfx}(\bfg_{ 2i },\bfg_{2j})$ is as good-path and due to Claim~\ref{claim:3}\eqref{claim:3_31},  for all $1 \leq t \leq \ell/2 -1$, it holds that  $\{\bfg_{s_{2t-1}},\bfg_{s_{2t}}\}_\mathbf{p}$ is a pair-type edge. Thus,   for all $1 \leq t \leq \ell/2 -1$
	$$\bfg'_{s_{2t-1}} + \bfg'_{s_{2t}} = \bfg_{s_{2t-1}}+ \bfx + \bfg_{s_{2t}} +  \bfx =  \bfg_{s_{2t-1}}+  \bfg_{s_{2t}}.$$
	Therefore, it is deduced that for all $p\in[n/2]\setminus\{i,j\}$, it holds that
	\begin{align*}
	\bfg'_{2p} + \bfg'_{2p+1} = \bfg_{2p}  + \bfg_{2p+1}.
	\end{align*}
	 In case that $p=i$ or $p=j$, by Claim~\ref{claim:3}\eqref{claim:3_1} the columns $\bfg_{2i+1}$ and $\bfg_{2j+1}$  are not on the path $P_{\bfx}(\bfg_{ 2i },\bfg_{2j})$. Thus, $\bfg'_{2i+1} = \bfg_{2i+1}$ and $\bfg'_{2j+1} = \bfg_{2j+1}$. Therefore,
	\begin{align*}
	\bfg'_{2p} + \bfg'_{2p+1} = \bfg_{2p} + \bfg_{2p+1} + \bfx .
	\end{align*}
	
\end{IEEEproof}

  {
 Before proceeding to the next section, the following \emph{FindShortPath}($G, \bfx, t,m$) function is presented. Let $G$ be an \HGG and $\bfG_{\bfx}(G)$ be its graph for some $\bfx \in \F^s_2$. Let $\{\bfg_{2t},\bfg_{2t+1}\}_\mathbf{p}$ be a pair-type edge in $\bfG_{\bfx}(G)$. Assume that there is another pair-type edge  $\{\bfg_{2m},\bfg_{2m+1}\}_\mathbf{p}$ in $\bfG_{\bfx}(G)$ such that $m>t$. The \emph{FindShortPath}($G, \bfa, t,m$) function will be used under the condition that there is a cycle $C_i \in \bfC_{\bfx}(G)$ such that both $\{\bfg_{2t},\bfg_{2t+1}\}_\mathbf{p}$ and $\{\bfg_{2m},\bfg_{2m+1}\}_\mathbf{p}$ are in $C_i$.
}
 \begin{algorithm}[h!]
 	\renewcommand{\thealgorithm}{}
 	\floatname{algorithm}{}
 	\caption{\emph{FindShortPath}($G, \bfx, t,m$)}
 	\label{alg:0}
 	\begin{algorithmic}[1]
 		\State $P_{\bfx}  \leftarrow$  the good-path $P_{\bfx}(\bfg_{2t},\bfg_{2t+1},G)$
 		\State $d_1 \leftarrow d_{P_{\bfx}}(\bfg_{2t+1},\bfg_{2m})$
 		\State $d_2 \leftarrow d_{P_{\bfx}}(\bfg_{2t+1},\bfg_{2m+1})$
 		\If {$d_1 < d_2$}
 		\State $j \leftarrow 2m$
 		\Else
 		\State  $j \leftarrow 2m+1$
 		\EndIf
 	\end{algorithmic}
 \end{algorithm} 

The \emph{FindShortPath}($G, \bfx, t,m$) function is presented since it will be used several times in this paper.

\subsection{The {FindGoodOrBadRequest}($G,t, \bfv$) function}

Let $G$ be an \HG, let $\bfv \in \F^s_2$, and let $t\in[n/2]$. Denote $\bfy = \bfg_{2t} + \bfg_{2t+1}$. In this section we will show the function called \emph{FindGoodOrBadRequest}($G,t, \bfv$). This function will be used by the first algorithm which will be presented in the next section. The task of this function is to update the sum of the pair $\bfg_{2t} ,\bfg_{2t+1}$ to either $\bfv$ or $\bfv + \bfy$. It also changes the sum of the last pair $\bfg_{2t} , \bfg_{2t+1}$, but, this pair is used as a ``redundancy pair", i.e., it is not important what the sum of this pair. Another important thing to mention, is that the algorithm   \emph{FindGoodOrBadRequest}($G,t, \bfv$) do not update the sum of the pairs on indices $2p$ and $2p+1$ for all $p\neq t$, even though these columns could be reordered.  The case $\bfg_{2t} + \bfg_{2t+1} = \bfv,\bfg_{2t} + \bfg_{2t+1} = \bfv + \bfy$ is called a good, bad case and $t$ will, won't be inserted in $\cB$, respectively. We now ready to present the function.
\begin{algorithm}[h!]
 	\renewcommand{\thealgorithm}{}
	\floatname{algorithm}{}
	\caption{\emph{FindGoodOrBadRequest}($G,t, \bfv$)}
	\label{alg:1_1}
	\begin{algorithmic}[1]
		\State $\bfy \leftarrow \bfg_{2t} + \bfg_{2t+1}$
		\If {$\bfv = \bfy$} \label{alg1:st0}
		\State Return $G$ and $\cB$ 
		\EndIf
		\State $\bfu \leftarrow \bfg_{2t+1} + \bfg_{n-2}$ \label{alg1:st6}
		\For{$p=1,2, 3$}\label{alg1:st3}
		\If {$p=1$}\label{alg1:st4}
		\State $\bfa \leftarrow \bfv + \bfy$\label{alg1:st5}
		\EndIf
		\If{$p=2$}\label{alg1:st7}
		\State  $\bfa \leftarrow \bfv + \bfy + \bfu$\label{alg1:st8}
		\State  Swap the columns $\bfg_{2t+1}$ and $\bfg_{n-2}$ in $G$\label{alg1:st9}
		\EndIf
		\If{$p=3$}\label{alg1:st10}
		\State  $\bfa \leftarrow \bfv + \bfu$\label{alg1:st11}
		\State  Swap the columns $\bfg_{2t}$ and $\bfg_{n-2}$ in $G$\label{alg1:st12} 
		\EndIf
		\State $P_{\bfa}  \leftarrow$  the good-path $P_{\bfa}(\bfg_{2t},\bfg_{2t+1},G)$\label{alg1:st13}
		\State $\mathbf{r} \leftarrow  \{\bfg_{n-2},\bfg_{n-1}\}_{\mathbf{p}}$\label{alg1:st14}
		\If {$\mathbf{r} \in P_{\bfa}$ } \label{alg1:st15}
		\State $j \leftarrow$ \emph{FindShortPath}($G, \bfa, t,n/2$)
		\State $G'\leftarrow \cF_{\bfa}(\bfg_{2t+1},\bfg_j)$\label{alg1:st17}
		\State Return $G'$ and $\cB'$ \label{alg1:st18}
		\EndIf
		\EndFor
		\State $G'\leftarrow  \cF_{\bfa}(\bfg_{2t},\bfg_{2t+1})$\label{alg1:st19}
		\State Swap the columns $\bfg'_{2t}$ and $\bfg'_{n-2}$ of $G'$\label{alg1:st20}
		\State $\cB' \leftarrow \cB\cup \{t\}$\label{alg1:st21}
		\State Return $G'$ and $\cB'$ 
	\end{algorithmic}
\end{algorithm}

\begin{figure}[t!] 
	\hfill
	\subfigure[The $p=1$ case, $\bfa = \bfv + \bfy$.]{\includegraphics[width=43.2mm]{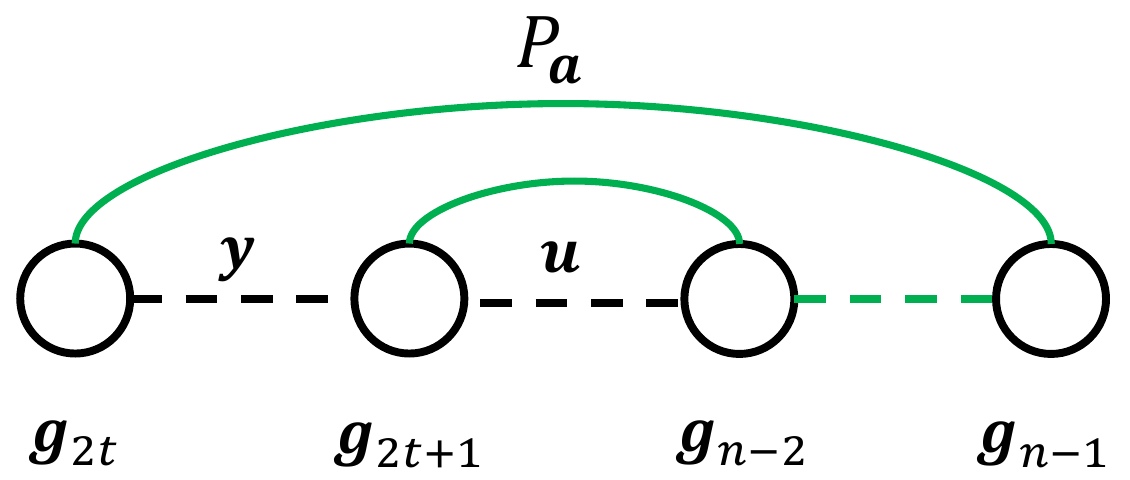}} 
	\hfill
	\subfigure[The $p=2$ case, $\bfa = \bfv + \bfy +\bfu$.]{\includegraphics[width=43.2mm]{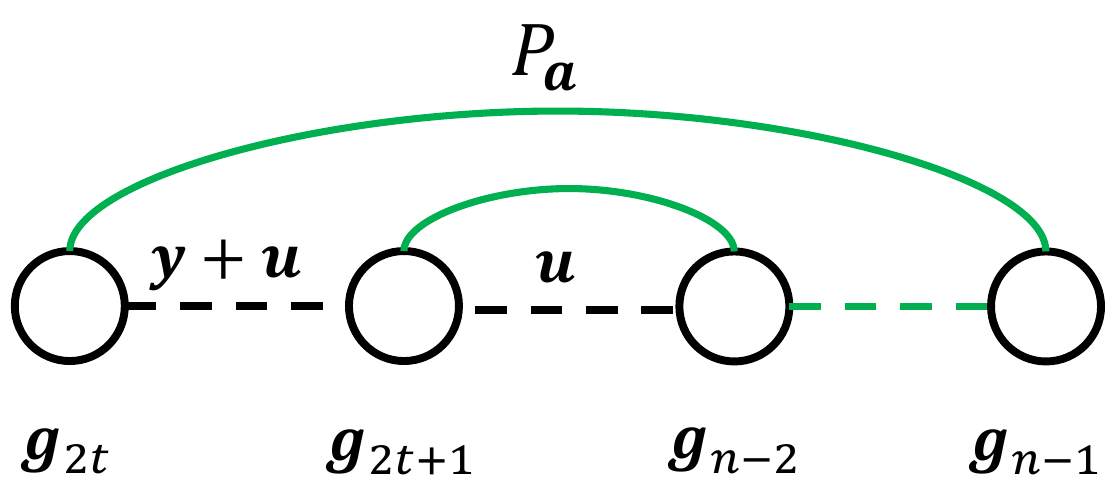}} 
	\hfill
	\subfigure[The $p=3$ case, $\bfa = \bfv + \bfu$.]{\includegraphics[width=43.2mm]{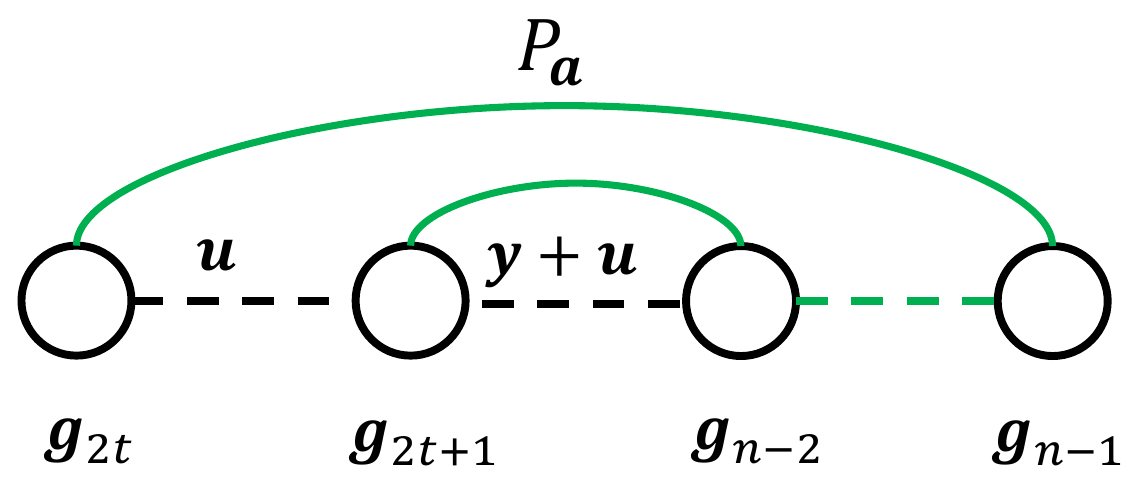}} 
	\hfill
	\subfigure[The bad case, $\bfa = \bfv + \bfu$.]{\includegraphics[width=43.2mm]{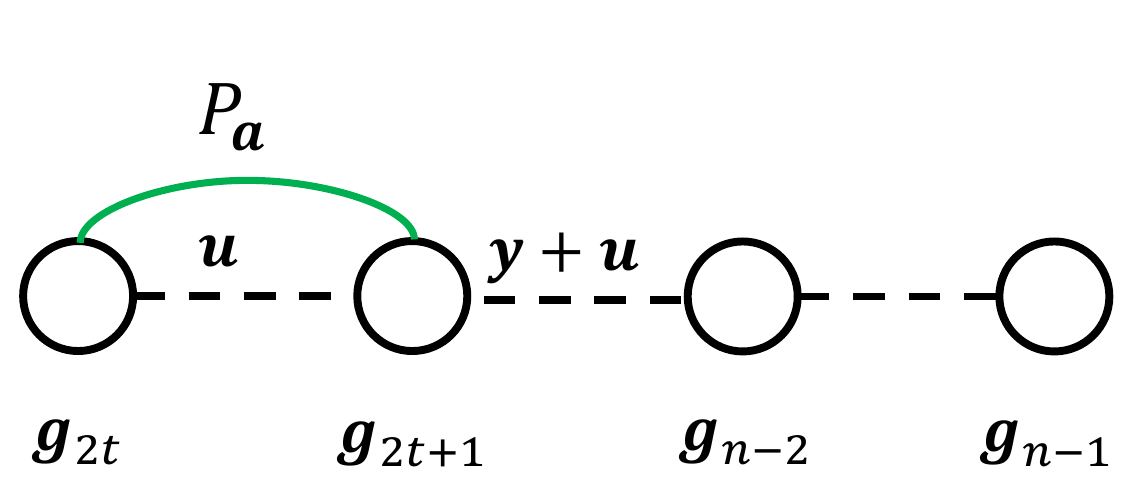}} 
	\hfill
	\caption{Explanation of the function \emph{FindGoodOrBadRequest}($G,t, \bfv$).}\label{fig3}
\end{figure}

An explanation of the \emph{FindGoodOrBadRequest}($G,t, \bfv$) function is shown in the next example.
\begin{example}
	In Fig~\ref{fig3} we illustrate three good situations in which Step~\ref{alg1:st15} in the function \emph{FindGoodOrBadRequest}($G,t, \bfv$) succeeds, and one bad case in which Step~\ref{alg1:st15} in the function \emph{FindGoodOrBadRequest}($G,t, \bfv$) fails. The solid green line in all figures is a sub-path of the good-path $P_{\bfa}$ (which is a path between the nodes $\bfg_{2t},\bfg_{2t+1}$ in $\bfG_{\bfa}(G)$). The dashed lines represent the edges between the signed nodes. The green dashed line is an edge on $P_{\bfa}$. Without loss of generality, it is assumed that the closest node between $\bfg_{n-2}$ and $\bfg_{n-1}$ to $\bfg_{2t+1}$ in $P_{\bfa}$ is $\bfg_{n-2}$.
	The labels of the edges represent the summation of the vectors of its incident nodes.
	Each of the three good cases illustrated in (a)-(c) lead to the fact that a pair $\bfg'_{2t},\bfg'_{2t+1}$ will be summed up to $\bfv$ (Step~\ref{alg1:st17}). In the bad case illustrated by (d), this pair will be summed up only to $\bfv + \bfy$ (Steps~\ref{alg1:st19}-\ref{alg1:st20}).
\end{example}

Denote by $\mathds{1} \in \{0,1\}$ a binary indicator  such that $\mathds{1} = 1$ if and only if  the function \emph{FindGoodOrBadRequest}($G,t, \bfv$) reaches Step~\ref{alg1:st19}. Our next goal is to prove the following important lemma. 
\begin{lemma}\label{lemma:11}
	The function {FindGoodOrBadRequest}($G,t, \bfv$) will generate a matrix 
	$$G'=  [\bfg'_0,\bfg'_1,\dots,\bfg'_{n-1}]$$ 
	such that
	\begin{align*}
	\bfg'_{2p} + \bfg'_{2p+1} =
	\begin{cases}
	\bfg_{2p} + \bfg_{2p+1}  &  p\neq t,n/2-1 \\
	\bfv + \mathds{1} \bfy   & p =t
	\end{cases}.
	\end{align*}
\end{lemma}

\begin{IEEEproof}
	First we show that  if the function reaches Step~\ref{alg1:st18}, then
	\begin{align}\label{eq:4}
	\bfg_{2t} + \bfg_{2t+1} +\bfa = \bfv.
	\end{align} 
	We separate the proof for the three cases of $p\in\{1,2,3\}$. To better understand these cases we refer the reader to Fig.~\ref{fig3}(a)-(c). Remember that by Step~\ref{alg1:st6}, $\bfu = \bfg_{2t+1} + \bfg_{n-2}.$
	\begin{enumerate}
		\item If $p=1$, then $\bfg_{2t} + \bfg_{2t+1} = \bfy$. By Step~\ref{alg1:st5}, $\bfa=\bfv + \bfy$, and therefore equality~\eqref{eq:4} holds. 
		\item If $p=2$, then by Step~\ref{alg1:st9}, after swapping $\bfg_{2t+1}$ and $\bfg_{n-2}$, it is deduced that 
		$$\bfg_{2t} + \bfg_{2t+1} = \bfy + \bfu.$$ 
		By Step~\ref{alg1:st8}, $\bfa = \bfv + \bfy + \bfu$, which concludes the correctness of equality~\eqref{eq:4}. 
		\item If $p=3$, then by Step~\ref{alg1:st12}, after swapping $\bfg_{2t}$ and $\bfg_{n-2}$, it is deduced that 
		$$\bfg_{2t} + \bfg_{2t+1}=\bfu.$$ 
		By Step~\ref{alg1:st11}, $\bfa = \bfv +\bfu$, corresponding to~\eqref{eq:4}. 
	\end{enumerate}
	
	Note that by Claim~\ref{claim:3_2}, there is always a good-path between $\bfg_{2t}$ and $\bfg_{2t+1}$.
	Now, suppose that in one of these $3$ cases, there is a good-path  $P_{\bfa}(\bfg_{2t},\bfg_{2t+1})$ in  $\bfG_{\bfa}(G)$ which includes the edge $\{\bfg_{n-2},\bfg_{n-1}\}_\mathbf{p}$ (with respect to Step~\ref{alg1:st15}). 
	 {By executing \emph{FindShortPath}($G, \bfa, t,n/2$)} we find the closest node between $\bfg_{n-2}$ and $\bfg_{n-1}$ to  $\bfg_{2t+1}$ on the path $P_{\bfa}$. This node is  denoted by $\bfg_j$. Thus, the first and the last edges on the path $P_{\bfa}(\bfg_{2t+1},\bfg_j)$ have to be $\bfx$-type edges. By definition, it is deduced that  $P_{\bfa}(\bfg_{2t+1},\bfg_j)$ is a good-path. According to Step~\ref{alg1:st17}, $G'= \cF_{\bfa}(\bfg_{2t+1},\bfg_j)$. By Lemma~\ref{lemma:8} this step changes the pair summations of only the pairs $\bfg_{2t},\bfg_{2t+1}$ and $\bfg_{n-2},\bfg_{n-1}$.
	More precisely, 
	$$\bfg'_{2t} + \bfg'_{2t+1} = \bfg_{2t} + \bfg_{2t+1} + \bfa = \bfv$$ and the sum of the pair $\bfg'_{n-2},\bfg'_{n-1}$ does not matter. 
	
	Finally, if the function does not succeed to find any of these good-paths, we will show that it will create a matrix $G'$ such that the pair $\bfg'_{2t},\bfg'_{2t+1}$ will be almost correct, that is,
	$$\bfg'_{2t} + \bfg'_{2t+1} = \bfv+\bfy.$$ 
	This will be done in Steps~\ref{alg1:st19}--\ref{alg1:st20}. First, the path $P_{\bfa}(\bfg_{2t},\bfg_{2t+1})$ does not include the edge $\{\bfg_{n-2},\bfg_{n-1}\}_\mathbf{p}$ since Step~\ref{alg1:st15} has failed.
	Second, the function makes two swaps in Step~\ref{alg1:st9} and Step~\ref{alg1:st12} such that 
	$$\bfg_{2t} + \bfg_{2t+1} = \bfu,$$
	and 
	$$\bfg_{2t+1}+ \bfg_{n-2} = \bfu+\bfy,$$
	and by Step~\ref{alg1:st11}, $\bfa = \bfv + \bfu$. This is illustrated in Fig.~\ref{fig3}(d). Thus, according to Step~\ref{alg1:st19}, $G' = \cF_{\bfa}(\bfg_{2t},\bfg_{2t+1})$ such that
	$$\bfg'_{2t} + \bfg'_{2t+1} = \bfg_{2t} + \bfg_{2t+1} = \bfu,$$
	and the column vectors $\bfg_{n-2},\bfg_{n-1}$ are not changed. Therefore,
	$$\bfg'_{2t+1}+ \bfg'_{n-2} = \bfg_{2t+1}+ \bfg_{n-2}+\bfa = \bfv+\bfy.$$
	By Step~\ref{alg1:st20} in which the columns  $\bfg'_{2t},\bfg'_{n-2} $ are swapped we get
	$$\bfg'_{2t} + \bfg'_{2t+1} = \bfv+\bfy.$$ 
	
	We conclude that the function will generate an \HG
	$$G'=  [\bfg'_0,\bfg'_1,\dots,\bfg'_{n-1}]$$ 
	such that
	\begin{align*}
	\bfg'_{2p} + \bfg'_{2p+1} =
	\begin{cases}
	\bfg_{2p} + \bfg_{2p+1}  &  p\neq t,n/2-1 \\
	\bfv + \mathds{1} \bfy   & p = t
	\end{cases}
	\end{align*}
	where $\mathds{1} = 1$ if and only if  the function reached Step~\ref{alg1:st19}.
	
\end{IEEEproof}

\subsection{The First Algorithm}

 
We start with the first algorithm which is referred by \emph{FBSolution}($\tau$,$M$), where $M$ is the request and $\tau$ will be the number of iterations in the algorithm, which is the number of columns in $M$. We define more variables that will be used in the routine of \emph{FBSolution}($\tau$,$M$), and some auxiliary results. 
The $\tau$ iterations in the algorithm operate as follows. First, we demand that the initial state of the matrix 
 $$G = [\bfg_0,\bfg_1,\dots,\bfg_{n-1}]$$ 
will satisfy
\begin{align}\label{eq:1}
&\bfg_{2t} + \bfg_{2t+1} = \bfe  , & t\in [n/2].
\end{align}
The matrix $G$ exists due to the following claim.
 
 \begin{claim}\label{claim:2}
 	There is an \HGG $G=[\bfg_0,\bfg_1,\dots,\bfg_{n-1}]$
 	such that for all $ t\in [n/2]$
	\begin{align*}
	&\bfg_{2t} + \bfg_{2t+1} = \bfe.
	\end{align*}
 \end{claim}
 
 \begin{IEEEproof}
 	Such an \HGG $G$ is constructed by taking an order of its column vectors such that  for all $t\in[n/2]$,
 	\begin{align*}
 	\bfg_{ 2t } = (z_0,z_1,\dots,z_{s-2},0),~~~	\bfg_{ 2t+1 } = (z_0,z_1,\dots,z_{s-2},1).
 	\end{align*}
 \end{IEEEproof}

  {The following corollary states that Claim~\ref{claim:2} holds for all $\bfx \in \F^s_2$ instead of $\bfe$. The proof of this corollary is similar to the one of Claim~\ref{claim:2}.}
\begin{corollary}\label{cor:1}
	For any $M$ that has one kind of request $\bfv_j$, there is a Hadamard solution for $M$.
\end{corollary}

  {Extending Corollary~\ref{cor:1} to the cases where there are at most a fixed number of different requests $d$ is an interesting problem by itself, which is out of the scope of this paper. For the case of $d=3$ we believe we have a proof, however it is omitted since we found it to be long and cumbersome. Finding a simple solution for this case and in general for arbitrary $d$ is left for future research.}

According to Corollary~\ref{cor:1}, for the rest of this section we assume that $M$ has at least two kinds of requests $\bfv_j$. It is also assumed that $\tau = \lfloor\frac{2}{3}\cdot 2^{s-1} \rfloor$. The \HGG at the end of the $t$-th iteration will be denoted by $$G^{(t+1)} = [\bfg^{(t+1)}_0,\bfg^{(t+1)}_1,\dots,\bfg^{(t+1)}_{n-1}].$$ 
Now we are ready to present the  \emph{FBSolution}($\tau$,$M$) algorithm.

\begin{algorithm}[h!]
	 \renewcommand{\thealgorithm}{}
	\floatname{algorithm}{Algorithm 1}
	\caption{\emph{FBSolution}($\tau, M$)}
	\label{alg:1}
	\begin{algorithmic}[1]
		\State  $G^{(0)} \leftarrow G$  
		\State $\cB^{(0)} = \emptyset$\label{alg1:st1}
		\For{$t=0,\ldots, \tau-1$}\label{alg1:st2}
		\State $G^{(t+1)}, \cB^{(t+1)}\leftarrow$ \emph{FindGoodOrBadRequest}($G^{(t)},t, \bfv_t$)
		\EndFor
		\State Return $G^{(\tau)}$ and $\cB^{(\tau)}$\label{alg1:st22}
	\end{algorithmic}
\end{algorithm}

At the end of the \emph{FBSolution}($\tau$,$M$) algorithm we obtained the set $\cB_1 = \cB^{(\tau)}$ and the matrix $G^{(\tau)} = [ \bfg^{(\tau)}_0,\bfg^{(\tau)}_1,\dots,\bfg^{(\tau)}_{n-1}]$. By Definition~\ref{def:2}, the set $\cB_1$ uniquely defines the triple-set $(\cG_1,\cB_1,\cR_1)$. Since in our case $\bfy = \bfe$, by Lemma~\ref{lemma:11}, for all $t\in [n/2]$, the matrix $G^{(\tau)}$ satisfies that
\begin{align*}
\bfg^{(\tau)}_{2t} + \bfg^{(\tau)}_{2t+1} =
\begin{cases}
\bfv_t & t\in \cG_1\\
\bfv_t + \bfe  & t\in \cB_1 \\
\bfe  & t\in \cR_1
\end{cases}.
\end{align*}
Let  $\cM_1 = [\bfw_0, \bfw_1, \dots,\bfw_{2^{s-1}-1}]$ be the matrix such that  for all $t\in [n/2]$
$$\bfw_t = \bfg^{(\tau)}_{2t} +  \bfg^{(\tau)}_{2t+1}.$$
Therefore, it is deduced that $\cM_1= \cM_1(\cG_1,\cB_1,\cR_1)$ is a triple-matrix  of $M$. 
By definition of $ \cM_1(\cG_1,\cB_1,\cR_1)$, the matrix $G^{(\tau)}$ is its Hadamard solution. If the set $\cB_1$ satisfies $|\cB_1|\leq\lfloor\frac{1}{3}\cdot~2^{s-1} \rfloor$ then by Lemma~\ref{lemma:5} there is a solution for $M$. Otherwise, we will make another reordering on the columns of $G^{(\tau)}$ in order to obtain a new bad set $\cB_2$ for which $|\cB_2|\leq\lfloor\frac{1}{3}\cdot 2^{s-1} \rfloor$. This will be done in the next section by showing our second algorithm.

	\subsection{The Second Algorithm}
	
	
	From now on we assume that 
	$$G^{(\tau)} = G = [\bfg_0,\bfg_1,\dots,\bfg_{n-1}]$$ 
	and $\cB=\cB_1$. Before showing the third second we start with the following definition.

	\begin{definition}
		Let $\bfC_\bfe(G)$ be a partition of simple cycles in $\bfG_\bfe(G)$. A pair of distinct indices $t_1,t_2 \in \cB$ is called a \textbf{bad-indices pair} in $\bfC_\bfe(G) $ if  both edges $\{\bfg_{2t_1},\bfg_{2t_1+1} \}_\mathbf{p}$ and  $\{\bfg_{2t_2},\bfg_{2t_2+1} \}_\mathbf{p}$ are in the same simple cycle in $\bfG_\bfe(G)$.
	\end{definition}
	Now we show the algorithm \emph{ClearBadCycles}($G,\cB$) in which the columns of the \HGG $G$ are reordered, and the set $\cB$ will be modified and its size will be decreased. 
\begin{algorithm}[h!]
	\renewcommand{\thealgorithm}{}
	\floatname{algorithm}{Algorithm 2}
	\caption{\emph{ClearBadCycles}($G,\cB$)}
	\label{alg:1.5}
	\begin{algorithmic}[1]
		\State $\bfC_{\bfe}(G)\leftarrow $ The partition of simple cycles in $\bfG_\bfe(G)$ \label{alg2:st1}
		\While {$\exists t_1,t_2$ a bad-indices pair in $\bfC_{\bfe}(G)$} \label{alg2:st2}
		\State $j \leftarrow$ \emph{FindShortPath}($G, \bfe, t_1,t_2$) \label{alg2:st3} 
		\State $G\leftarrow\cF_{\bfe}(\bfg_{2t_1+1},\bfg_j)$ \label{alg2:st6}
		\State $\bfC_{\bfe}(G)\leftarrow $ The partition of simple cycles in $\bfG_\bfe(G)$ \label{alg2:st7}
		\State Remove $t_1,t_2$ from $\cB$\label{alg2:st12}
		\EndWhile
		\State Return $G$ and $\cB$  
	\end{algorithmic}
\end{algorithm}  

Let $ G_2 = [\bfg_0^\star,\bfg_1^\star,\dots,\bfg_{n-1}^\star]$  be the \HGG and $\cB_2$ be the bad set output of the \emph{ClearBadCycles}($G,\cB$) algorithm.
We remind the reader that $M = [\bfv_0, \bfv_1, \dots,\bfv_{\lfloor\frac{2}{3}\cdot 2^{s-1} \rfloor-1} ]$.
Let  $\cM_2 = [\bfw_0, \bfw_1, \dots,\bfw_{2^{s-1}-1}]$ be the matrix such that  for all $t\in [n/2]$
$$\bfw_t = \bfg^{\star}_{2t} +  \bfg^{\star}_{2t+1}.$$
Since $\cB_2$ uniquely defines the triple set $(\cG_2,\cB_2,\cR_2)$, it is deduced that the matrix $\cM_2$  is a triple-matrix $\cM_2(\cG_2,\cB_2,\cR_2) $ of $M$. Next, it will be shown that the algorithm \emph{ClearBadCycles}($G,\cB$) will stop and $|\cB_2|$ will be bounded from above by $\lfloor\frac{1}{3}\cdot 2^{s-1} \rfloor$ after the execution of the algorithm.
	\begin{lemma}\label{lem:10}
		The algorithm ClearBadCycles($G,\cB$) outputs a set $\cB_2$ such that $|\cB_2| \leq \lfloor\frac{1}{3}\cdot 2^{s-1} \rfloor$.
	\end{lemma}
	\begin{IEEEproof}
	According to Step~\ref{alg2:st2}  if there is a simple cycle containing a bad-indices pair $t_1,t_2$ in $\bfC_{\bfe}(G)$, the algorithm will enter the routine.
	Thus, there is a good-path between $\bfg_{2t_1+1}$ and one of the nodes $\bfg_{2t_2},\bfg_{2t_2+1}$ (the closest one between them to $\bfg_{2t_1+1}$), and the index of this node is denoted by $j$ (Step~\ref{alg2:st3}). 
	Since $t_1,t_2 \in \cB$, before the algorithm reaches Step~\ref{alg2:st6}, it holds
	$$ \bfg_{2t_1} +\bfg_{2t_1+1} = \bfv_{t_1}+\bfe ,~~ \bfg_{2t_2} +\bfg_{2t_2+1} = \bfv_{t_2}+\bfe. $$
	By executing $\cF_{\bfe}( \bfg_{2t_1+1},\bfg_j)$, due to Lemma~\ref{lemma:8}, the matrix $G$ is updated to a matrix $G'$ such that only the two following pair summations are correctly changed to
	$$ \bfg'_{2t_1} +\bfg'_{2t_1+1} = \bfv_{t_1} ,~~ \bfg'_{2t_2} +\bfg'_{2t_2+1} = \bfv_{t_2}. $$
	Thus, the indices $t_1$ and $t_2$ are removed from $\cB$ (Step~\ref{alg2:st12}). Therefore, Step~\ref{alg2:st2} will fail when each simple cycle will have at most one $t\in \cB$ such that   
	$$  \bfg_{2t} +\bfg_{2t+1} = \bfv_t + \bfe,$$
	and we will call it a ``bad cycle". Suppose that there are $p$ bad cycles at the end of the algorithm. We are left with showing that  $p \leq \lfloor\frac{1}{3}\cdot 2^{s-1} \rfloor$.
	
	Observe that for all $t\in \cR$, the nodes $\bfg_{2t}$ and $\bfg_{2t+1}$ are connected by two parallel edges, and therefore they create cycles of length $2$. These cycles are not bad cycles by definition. Since there are $2|\cR|$ such columns in $G$ and together with the pair $\bfg_{n-2},\bfg_{n-1}$ and $|\cR| = 2^{s-1}-\Big\lfloor\frac{2}{3}\cdot 2^{s-1}\Big\rfloor-1$, only the first  
	$$2\cdot |\cB\cup \cG| = 2\cdot \Big\lfloor\frac{2}{3}\cdot 2^{s-1} \Big\rfloor$$ 
	columns of $G$ can be partitioned into bad cycles. Our next goal is to prove that the size of each bad cycle is  at least $4$. 
	Assume to the contrary that there is a bad cycle of length $2$. 
	Since we are using the graph $\bfG_\bfe(G)$, such a simple cycle of two nodes $\bfg_{2t},\bfg_{2t+1}$, $t\in\cB$, satisfies that
	$$\bfg_{2t} +\bfg_{2t+1} = \bfe.$$
	In that case $\bfg_{2t} +\bfg_{2t+1} \neq  \bfv_t + \bfe$ since $\bfv_t$ is non-zero vector, so  $\bfg_{2t} +\bfg_{2t+1} = \bfv_t= \bfe$. According to Step~\ref{alg1:st1} in the function \emph{FindGoodOrBadRequest}($G,t, \bfv$),
	$t\notin \cB$, which results with a contradiction. Therefore, indeed all simple cycles are of size at least $4$. Thus, 
	$$|\cB| \leq p \leq \Big\lfloor \frac{1}{4}\cdot \Big(2 \cdot\Big\lfloor\frac{2}{3}\cdot 2^{s-1} \Big\rfloor \Big) \Big\rfloor= \Big\lfloor\frac{1}{3} 2^{s-1} \Big\rfloor,$$
	where the last equality holds since by the nested division $\left\lfloor  \frac{  \lfloor x/y \rfloor}{z} \right\rfloor = \lfloor  \frac{ x}{yz} \rfloor $ for real $x,y$ and a positive integer $z$.
	\end{IEEEproof}
	
	We are finally ready to prove the main result of this section. 	
	\begin{theorem}\label{theo:32}
		An $FB\bs(2^s-1,s,\lfloor\frac{2}{3}\cdot 2^{s-1} \rfloor)$ code exists.
	\end{theorem}
	\begin{IEEEproof}
	Using the result of Lemma~\ref{lem:10} it is deduced that the algorithm \emph{ClearBadCycles}($G,\cB$) outputs the set $\cB_2$ such that it size is at most $\lfloor\frac{1}{3}\cdot 2^{s-1}  \rfloor$. The \HGG $G_2$ is again a Hadamard solution for a triple-matrix $ \cM_2(\cG_2,\cB_2,\cR_2)$ of $M$. Thus, by using Lemma~\ref{lemma:5}, it is deduced that there is a solution for $M$.  After removing the all-zero column vector from $G$, the proof of this theorem is immediately deduced.
	\end{IEEEproof}

\section{A Construction of $FB\bs(2^s + \lceil(3\alpha-2)\cdot2^{s-2}\rceil -1 ,s,\lfloor\alpha\cdot 2^{s-1}\rfloor)$ Codes}\label{sec:cons3}
In this section we show how to construct  $FB\bs(2^s + \lceil(3\alpha-2)\cdot2^{s-2}\rceil -1 ,s,\lfloor\alpha\cdot 2^{s-1}\rfloor)$ codes where $2/3 \leq  \alpha \leq 1$. For convenience, throughout this section let $n=2^s$ and $m = 2^s +\lceil (3\alpha-2)\cdot2^{s-2} \rceil$. Note that since $\alpha\geq 2/3$ it holds that $m\geq n$. Let $\bfe = (0,0,\dots,0,1) \in \F^s_2$. 

The following two definitions extend \HGs from Definition~\ref{HG_matrices} and triple-matrices from Definition~\ref{def:2}. 
\begin{definition}
	A matrix  
	$G = [\bfg_0,\bfg_1,\dots,\bfg_{m-1}]$
	of order $s\times m$ over $\F_2$ is called an \textit{extended-\HG} if the matrix
	$H_G =  [\bfg_0,\bfg_1,\dots,\bfg_{n-1}]$
	is an \HGG of order $s\times n$ and for all $n \leq i \leq m-1$ it holds $\bfg_i = \bfe$. The \HGG $H_G$ will be called the \textit{$H$-part} of $G$.
\end{definition}

\begin{definition}\label{def:3}
	Three sets $\cG,\cB,\cR\subseteq [2^{s-1}]$ are called an $\alpha$-\textbf{triple-set}, and are denoted by $\alpha$-$(\cG,\cB,\cR)$, if the following properties hold
	\begin{align*}
	&\cG \subseteq \Big[\Big\lfloor\alpha\cdot 2^{s-1} \Big\rfloor\Big],\\
	&\cB =  \Big[\Big\lfloor\alpha\cdot 2^{s-1} \Big\rfloor\Big] \setminus \cG,   \\
	&\cR = [2^{s-1}]\setminus \Big(\cG\cup \cB\cup\{2^{s-1}-1\}\Big).
	\end{align*}
	Given a matrix
	$M = [\bfv_0, \bfv_1, \dots,\bfv_{\lfloor\alpha\cdot 2^{s-1} \rfloor-1}]$
	of order ${s\times \lfloor\alpha\cdot 2^{s-1} \rfloor}$, a matrix $\cM(\cG,\cB,\cR) = [\bfw_0, \bfw_1, \dots,\bfw_{2^{s-1}-1}]$ of order  $s\times 2^{s-1}$ is referred as an $\alpha$-\textbf{triple-matrix of} $M$ if it holds that
	\begin{align*}
	\bfw_t =
	\begin{cases}
	\bfv_t & t\in \cG\\
	\bfv_t + \bfe  & t\in \cB \\
	\bfe  & t\in \cR
	\end{cases}.
	\end{align*}
\end{definition}

Note that by taking $\alpha = 2/3$, Definition~\ref{def:3} will be equivalent to Definition~\ref{def:2}.
Denote $\cN = [m]\setminus[n]$ and note that $|\cN| = \lceil(3\alpha-2)\cdot2^{s-2} \rceil $. We seek to design an algorithm which is very similar to the \emph{FBSolution}($\tau, M$) algorithm in the following respect. This algorithm will output an \HGG $H_G$ which will be the $H$-part of an extended-\HGG $G = [\bfg_0,\bfg_1,\dots,\bfg_{m-1}]$, and a set $\cB$. The set $\cB$ will define uniquely the $\alpha$-triple-set $\alpha$-$(\cG,\cB,\cR)$. For all $t\in \cG$ we will get
\begin{align*}
\bfg_{2t} + \bfg_{2t+1} = \bfv_t,
\end{align*}
and for all $t\in \cB$ we will get an almost desirable solution, that is,
\begin{align*}
\bfg_{2t} + \bfg_{2t+1} = \bfv_t + \bfe.
\end{align*}
The summation of the last pair $\bfg_{n-2},\bfg_{n-1}$ will be arbitrary.
Similarly to the technique that was shown in Section~\ref{sec:cons1}, the set $\cR$ will be used to correct the summations $\bfv_t +\bfe$ to $\bfv_t$, where $t\in\cB$. For that, the \emph{ClearBadCycles}($G,\cB$) algorithm will be used as it was done in Section~\ref{sec:cons1} in order to obtain an extended-\HGG $G$ and a set $\cB$ such that  $|\cB| \leq  \lfloor\frac{1}{2}\alpha\cdot 2^{s-1}  \rfloor$.
Even though $|\cB| \leq  \lfloor\frac{1}{2}\alpha\cdot 2^{s-1}  \rfloor$, the size of $\cR$ will not be bigger than the size of $\cB$ for $\alpha>2/3$. Thus, in this case, not all bad summation can be corrected. For that, we define the set $\cN$ that is also used to correct the summations $\bfv_t +\bfe$ to $\bfv_t$, where $t\in \cB$. This will be done based on the property that for all $t\in \cN$ it holds that $\bfg_t = \bfe$. 
In case that $\alpha <1$, together with $\cR$ and $\cN$, the last pair $\bfg_{n-2},\bfg_{n-1}$ will be used for the correction of these summations. Thus, if the inequality $|\cB| \leq |\cR| + |\cN| + 1$ holds, then it is possible to construct a solution for $M$. In case that $\alpha = 1$, we obtain  $|\cR| =0$. In this case, we will show how to get the inequality  $|\cB| \leq |\cN|$, which will similarly lead to a solution for $M$. Even though the last pair $\bfg_{n-2},\bfg_{n-1}$ has an arbitrary summation, it will still be shown how to obtain the request $\bfv_{n/2-1}$ from this pair. Therefore, our first goal is to show a condition which assures that either $|\cB| \leq |\cR| +|\cN| + 1$ or $|\cB| \leq |\cN|$. This is done in Claim~\ref{claim:6}.

\begin{claim}\label{claim:6}
	Let  $(\cG,\cB,\cR)$ be an $\alpha$-triple-set where ${|\cB| \leq  \lfloor\frac{1}{2}\alpha\cdot 2^{s-1}  \rfloor}$. If $2/3 \leq \alpha<1$,  then $|\cB| \leq |\cR| +|\cN| + 1$, and if $\alpha =1$ then  $|\cB| \leq |\cN|$. 
\end{claim}

\begin{IEEEproof}
	Let $2/3 \leq \alpha <1$.
	According to the definition of $\alpha$-$(\cG,\cB,\cR)$, since $\cG\cup \cB = [\lfloor \alpha\cdot 2^{s-1} \rfloor]$ it holds that 
	\begin{align*}
	|\cR| &= \Big|[2^{s-1}]\setminus \Big(\cG\cup \cB\cup\{2^{s-1}-1\}\Big)\Big|\\
	&= 2^{s-1} -  \Big\lfloor\alpha\cdot 2^{s-1} \Big\rfloor -1.
	\end{align*}
	We also know that $|\cN| = \lceil(3\alpha-2)\cdot2^{s-2} \rceil $.	Thus, in order to prove that $|\cB| \leq |\cR| + |\cN|+1$, since  $|\cB| \leq  \lfloor\frac{1}{2}\alpha\cdot 2^{s-1} \Big\rfloor$, it is enough to prove inequality $(a)$ in 
	\begin{align*}
	|\cR| + |\cN| + 1 =& 2^{s-1} -  \Big\lfloor \alpha\cdot 2^{s-1} \Big\rfloor   +  \Big\lceil (3\alpha-2) \cdot 2^{s-2} \Big\rceil\\
	\stackrel{(a)}{\geq}& \Big\lfloor\frac{1}{2}\alpha\cdot 2^{s-1} \Big\rfloor \geq |\cB|.
	\end{align*}
	Inequality $(a)$ is equivalent to
	\begin{align*}
	2^{s-1}   \geq    \Big\lfloor 2\alpha\cdot 2^{s-2} \Big\rfloor   + \Big\lfloor \alpha\cdot 2^{s-2} \Big\rfloor  - \Big\lceil(3\alpha-2)\cdot2^{s-2} \Big\rceil,
	\end{align*}
	which holds since
	\begin{align*}
	& \Big\lfloor 2\alpha\cdot 2^{s-2} \Big\rfloor   + \Big\lfloor \alpha\cdot 2^{s-2} \Big\rfloor  - \Big\lceil(3\alpha-2)\cdot2^{s-2} \Big\rceil \\
	&\leq    2\alpha\cdot 2^{s-2}    +  \alpha\cdot 2^{s-2}    -  (3\alpha-2)\cdot2^{s-2}\\
	& =  2^{s-2}(2\alpha     +  \alpha  -  (3\alpha-2) )  = 2\cdot 2^{s-2} = 2^{s-1}.
	\end{align*}
	Now  if $\alpha =1$, then $|\cB| \leq  \lfloor\frac{1}{2}\cdot 2^{s-1}  \rfloor = 2^{s-2}$. By the definition of $\cR$, it holds that $|\cR| = 0$ and  by the definition of $\cN$ it holds that $|\cN| = 2^{s-2}$. Therefore $|\cB| \leq  2^{s-2} = |\cN|$.
\end{IEEEproof}

Let $M = [\bfv_0, \bfv_1, \dots,\bfv_{ \lfloor \alpha\cdot 2^{s-1} \rfloor-1}]$ be a request of order $s\times \lfloor  \alpha  \cdot2^{s-1} \rfloor$. Our goal is to construct an extended-\HGG of order $s\times m$ which will provide a solution for $M$. For that, the $\alpha$-\emph{FBSolution}($M$) algorithm is presented. In this algorithm, the matrix $H$ is represented by $H = [\bfg_0,\bfg_{1},\dots,\bfg_{n-1}]$.

\begin{algorithm}[h!]
	\renewcommand{\thealgorithm}{}
	\floatname{algorithm}{Algorithm 3}
	\caption{$\alpha$-\emph{FBSolution}($M$)}
	\label{alg:3}
	\begin{algorithmic}[1]
		\If {$\alpha < 1$}
		\State $\tau \leftarrow \lfloor  \alpha  \cdot2^{s-1} \rfloor $
		\ElsIf {$\alpha = 1$}
		\State  $\tau \leftarrow  2^{s-1}-1$
		\EndIf
		\State  $H,\cB \leftarrow $ \emph{FBSolution}($\tau$,$M$)\label{alg:3st:5}
		\State $H,\cB \leftarrow$ \emph{ClearBadCycles}($H$,$\cB$)\label{alg:3st:6}
		\If {$\alpha<1$ and $\bfg_{n-2}\neq \bfe$}\label{alg:3st:7}
		\State $H\leftarrow  \cF_{\bfg_{n-2} + \bfe}(\bfg_{n-2},\bfg_{n-1})$\label{alg:3st:8}
		\EndIf
		\If{$\alpha=1$ and $\bfg_{n-2}\neq \bfv_{n/2-1}$}\label{alg:3st:9}
		\State $H\leftarrow  \cF_{\bfg_{n-2} + \bfv_{n/2-1}}(\bfg_{n-2},\bfg_{n-1})$\label{alg:3st:10}
		\EndIf
		
		\State Return $H$ and $\cB$  \label{alg:3st:11}
	\end{algorithmic}
\end{algorithm} 

Denote by $G = [\bfg_0,\bfg_1,\dots,\bfg_{m-1}]$ an extended-\HGG of order $s\times m$ such that the output matrix $H$ from the $\alpha$-\emph{FBSolution}($M$) algorithm is its $H$-part, i.e., $H_G = H$. Note that Steps~\ref{alg:3st:5}--\ref{alg:3st:6} define the set $\cB$. This set is obtained using a similar technique to the one from Section~\ref{sec:cons1}, except to the fact that here $2/3 \leq \alpha \leq 1$, while in Section~\ref{sec:cons1}, $\alpha = 2/3$. 
It is important to note that the size of $\cB$ is bounded due to the execution of the \emph{ClearBadCycles}($H$,$\cB$) algorithm (Step~\ref{alg:3st:6}). Therefore, we only state the following lemma since its proof is very similar to the one that was shown in Lemma~\ref{lem:10}. 
\begin{lemma}\label{lem:20}
	The $\alpha$-FBSolution($M$) algorithm outputs a set $\cB$ such that $|\cB| \leq \lfloor\frac{1}{2}\alpha\cdot~2^{s-1} \rfloor$.
\end{lemma}

We will use Lemma~\ref{lem:20} while proving the main theorem of this section.
\begin{theorem}\label{theo:16}
	For any $2/3 \leq \alpha \leq 1$, a functional batch code
	$$FB\bs(2^s +\lceil (3\alpha-2)\cdot2^{s-2}\rceil -1 ,s,\lfloor \alpha\cdot 2^{s-1} \rfloor)$$ exists.
\end{theorem}

\begin{IEEEproof}
	After finishing the $\alpha$-\emph{FBSolution}($M$) algorithm, we obtain an
	\HGG $H$ which is the $H$-part of the extended-\HGG 
	$ G = [\bfg_0,\bfg_1,\dots, \bfg_m].$ Remember that by the definition of $G$ and $\cN$, for all $t\in \cN$, it holds that $\bfg_t = \bfe$.
	In Step~\ref{alg:3st:5} we invoke the algorithm \emph{FBSolution}($\tau$,$M$). Therefore, for all $t< \tau$, there exists $ \mathds{1}_t \in \{0,1\}$ such that 
	$$\bfg_{2t} + \bfg_{2t+1} = \bfv_t +  \mathds{1}_t\bfe.$$ 
	Let $\alpha$-$(\cG,\cB,\cR)$ be an $\alpha$-triple-set that is uniquely defined by $\cB$ according to Definition~\ref{def:3}.
	Clearly,  for all $t\in \cG$, the recovery set $R_t$ is $R_t = \{2t,2t+1\}$.
	By Lemma~\ref{lem:20} it holds that  $|\cB| \leq \lfloor\frac{1}{2}\alpha\cdot~2^{s-1} \rfloor$.  We separate this proof for two cases.\\ 
	
	\textbf{Case 1:} Assume that $\alpha<1$. Due to  Lemma~\ref{lem:20} and Claim~\ref{claim:6}, if $\alpha<1$, it is deduced that $|\cB| \leq |\cR| +|\cN|+1 $. Let $\mathbf{t}$ be the maximum number in $\cB$ and let $\cB' = \cB\setminus \{\mathbf{t}\}$. Thus, $|\cB'| \leq |\cR| +|\cN|$.
	Therefore,  for all $t\in \cB'$, $R_t$ will have a different $t'$ such that $R_t$ equals to either $\{2t,2t+1,2t',2t'+1\}$ where $t'\in \cR$, or $\{2t,2t+1,t'\}$ where $t'\in \cN$. Thus, we showed the recovery sets for all requests except of $\bfv_{\mathbf{t}}$. Remember that  $\bfg_{2\mathbf{t}} + \bfg_{2\mathbf{t}+1} = \bfv_{\mathbf{t}} + \bfe$, and note that if $\bfg_{n-2} = \bfe$, this case is finished. Otherwise, $\bfg_{n-2} \neq \bfe$.
	This is handled by Steps~\ref{alg:3st:7}--\ref{alg:3st:8} as follows. By Claim~\ref{claim:3_2}, since $\bfg_{n-2}$ and $\bfg_{n-1}$ is a pair, we know that there is a good-path $P_\bfx(\bfg_{n-2},\bfg_{n-1})$ in an $\bfx$-type graph  $\bfG_{\bfx}(G)$ for all $\bfx \in \F^{s}_2$. Thus, if $\bfg_{n-2} \neq \bfe$, by taking $\bfx = \bfg_{n-2} + \bfe$, the algorithm can use the reordering function  $\cF_{\bfg_{n-2} + \bfe}(\bfg_{n-2},\bfg_{n-1})$, as it is done in Step~\ref{alg:3st:8}. By Lemma~\ref{lemma:8}, we obtain two new column vectors $\bfg'_{n-2}$ and $\bfg'_{n-1}$ such that
	\begin{align*}
	\bfg'_{n-2} = &\bfg_{n-2} + \bfg_{n-2} + \bfe = \bfe, \\
	\bfg'_{n-1} = &\bfg_{n-1} + \bfg_{n-2}+\bfe,
	\end{align*}
	without changing the summations of all other pairs on this path.
	Therefore, the recovery set for $\bfv_{\mathbf{t}}$ will be $R_{\mathbf{t}} = \{2\mathbf{t}, 2\mathbf{t}+1, n-2\}$, which concludes this case.\\
	
	\textbf{Case 2:} Assume that $\alpha = 1$.  Due to  Lemma~\ref{lem:20} and Claim~\ref{claim:6} if $\alpha = 1$ then $|\cB| \leq |\cN|$. Thus, similarly to Case 1, for all $t\in \cB$, the recovery sets $R_t$ can be obtained. However, we do not have a recovery set for $\bfv_{n/2-1}$ since the sum of the pair $\bfg_{n-2},\bfg_{n-1}$ is arbitrary. If $\bfg_{n-2} = \bfv_{n/2-1}$, then $R_{n/2-1} = \{n-2\}$. Otherwise, as in Case 1, by Step~\ref{alg:3st:10} it is deduced that $\bfg'_{n-2} = \bfv_{n/2-1}$.  Again $R_{n/2-1} = \{n-2\}$, which concludes this case.
	
	In both cases, after removing the all-zero column from $G$, we conclude the proof.

\end{IEEEproof}

\section{A Construction of $FB\bs(2^{s+1}-2,s,2^s)$  Codes}\label{sec:cons2}
In this section, a construction for  $FB\bs(2^{s+1}-2,s,2^s)$  codes will be shown by using the algorithm \emph{FBSolution}($\tau$,$M$). 
Throughout this section let $n=2^{s+1}$ and let $\bfe = (0,0,\dots,0,1) \in \F^{s+1}_2$. We start with the following definition.

\begin{definition}
	A matrix  $G = [\bfg_0,\bfg_1,\dots,\bfg_{2^{s+1}-1}]$ 
	of order $s\times 2^{s+1}$ over $\F_2$ such that each vector of $\F^s_2$ appears as a column vector in $G$ exactly twice, is called a \textit{double-\HG}. 
\end{definition}

Note that by removing the last row from any \HGG of order $(s+1)\times n$, we get a double-\HGG of order $s\times n$.
Also, note that each double-\HGG has exactly two all-zero columns. These columns will be removed at the end of the procedure, obtaining only 
$2^{s+1}-2$ column vectors. Next, the definition of a Hadamard solution is extended with respect to Definition~\ref{def:6.1}.

\begin{definition}\label{def:6.2}
	Let $ M  = [\bfv_0, \bfv_1, \dots,\bfv_{2^s-1} ]$ be a request of order $s\times 2^s$. The matrix $M$ has a \textbf{Hadamard solution} if there  exists a  double-\HGG  	$G = [\bfg_0,\bfg_1,\dots,\bfg_{2^{s+1}-1}]$ of order $s\times n$ such  that for all  $t\in [2^s-1]$,
	\begin{align*}
	&\bfv_t = \bfg_{ 2t } + \bfg_{2t+1},
	\end{align*}
	and for $t=n/2-1$ either $\bfv_t =  \bfg_{ n-2 } + \bfg_{n-1}$, or $\bfv_t = \bfg_{n-2}$, or $\bfv_t = \bfg_{n-1}$.
\end{definition}

Let $M = [\bfv_0, \bfv_1, \dots,\bfv_{2^s-1}]$ be a request of order $s\times 2^s$. Our goal is to construct a double-\HGG of order $s\times 2^{s+1}$ which will provide a Hadamard solution for $M$. 
Let 
$\cM = [\bfw_0, \bfw_1, \dots,\bfw_{2^s-1}]$ 
be a new matrix of order $(s+1)\times2^s$ generated by adding the all-zero row to $M$. 
Let $\tau = n/2-1$ and $\bf{0}_\ell$ be the all-zero vector of length $\ell$. 
We now show the algorithm \emph{OptFBSolution}($\cM$), which receives as an input the matrix $\cM$ and outputs a double-\HGG $G$ that will be a solution for $M$. As mentioned in the Introduction the returned solution is optimal.

\begin{algorithm}[h!]
	 \renewcommand{\thealgorithm}{}
	\floatname{algorithm}{Algorithm 4}
	\caption{\emph{OptFBSolution}($\cM$)}
	\label{alg:2}
	\begin{algorithmic}[1]
		\State $\tau \leftarrow  n/2-1$\label{alg:2st:1}
		\State  $G \leftarrow $ \emph{FBSolution}($\tau$,$\cM$)\label{alg:2st:2}
		\State $\bfy \leftarrow \sum^{2^s-1}_{i=0} \bfw_i$\label{alg:2st:3}
		\If{$\bfy \neq {\bf{0}}_{s+1}$}\label{alg:2st:4}
		\State $G\leftarrow  \cF_{\bfg_{n-2} + \bfy}(\bfg_{n-2},\bfg_{n-1})$\label{alg:2st:5}
		\EndIf		
		\State Remove the last row from $G$\label{alg:2st:6}
	\end{algorithmic}
\end{algorithm}  
The following lemma proves the correctness of Algorithm \emph{OptFBSolution}($\cM$).
 \begin{lemma}\label{lem:22}
 	The algorithm {OptFBSolution}($\cM$) outputs a double-\HGG $G'$ which is a Hadamard solution for $M$.
 \end{lemma}

\begin{IEEEproof}
	According to Step~\ref{alg:2st:2}, the algorithm \emph{FBSolution}($\tau$,$\cM$) is used with $\tau = n/2-1$. Thus, by Lemma~\ref{lemma:11} we obtain an \HGG 
	$  G=  [\bfg_0,\bfg_1,\dots,\bfg_{n-1}]$ 
	such that  for all $t\in [n/2-1]$
	\begin{align}\label{eq:4.1}
	\bfg_{2t} + \bfg_{2t+1} = \bfw_t + \mathds{1}_t\bfe,
	\end{align}
	where $\mathds{1}_t \in \{0,1\}$. 
	Let $G' = [\bfg'_0,\bfg'_1,\dots,\bfg'_{n-1}]$ be a double-\HGG of order $s\times n$ generated by removing the last row from $G$ according to Step~\ref{alg:2st:6}.
	Since $\cM = [\bfw_0,\bfw_1,\dots,\bfw_{2^s-1}]$ is generated by adding the all-zero row to $M$,
	by removing the last row from $G$ before Step~\ref{alg:2st:3},  for all $t\in [n/2-1]$ we could obtain $G'$ such that 
	\begin{align}\label{eq:4.2}
	\bfg'_{2t} + \bfg'_{2t+1} = \bfv_t.
	\end{align}
	However, $G'$ would provide a solution for $M$ except for the last request $\bfv_{n/2-1}$. We handle the last request using Steps~\ref{alg:2st:3}--\ref{alg:2st:5} that will be explained as follows. 

	Assume that $\sum^{2^s-1}_{i=0} \bfw_i =   \bfy $ and note that 
	\begin{align}\label{eq:5}
	\sum^{2^{s+1}-1}_{i=0} \bfg_i = {\bf{0}}_{s+1}.
	\end{align}
	Denote
	\begin{align}\label{eq:6}
	\mathds{1}_{n/2-1} = \sum^{2^{s}-2}_{t=0}\mathds{1}_t (\bmod 2).
	\end{align}
	 Thus, it is deduced that
	\begin{align*}
	\bfg_{n-2} + \bfg_{n-1}& \stackrel{(a)}{=} \sum^{2^{s+1}-3}_{i=0} \bfg_i \stackrel{(b)}{=} \sum^{2^s-2}_{t=0} (\bfw_t+ \mathds{1}_t\bfe)  \\
	&\stackrel{(c)}{=} \bfw_{n/2-1}+ \bfy + \mathds{1}_{n/2-1}\bfe.
	\end{align*}
	Equality $(a)$ holds due to~\eqref{eq:5}, equality $(b)$ holds according to~\eqref{eq:4.1}, and  equality $(c)$ holds by the definition of $\bfy$ and by~\eqref{eq:6}.
	Now if $\bfy =  {\bf{0}}_{s+1}$, according to Step 6, after removing the last row from $G$, we get $G'$ such that equation~\eqref{eq:4.2} holds also for $t=n/2-1$, that is,
	$$\bfg'_{n-2} + \bfg'_{n-1} = \bfv_{n/2-1}.$$
	Clearly, in this case $G'$ is a Hadamard solution for $M$.
	Otherwise, if  $\bfy \neq  {\bf{0}}_{s+1}$ then the algorithm enters the if condition in Step~\ref{alg:2st:4}. 
	By Claim~\ref{claim:3_2}, since $\bfg_{n-2}$ and $\bfg_{n-1}$ is a pair, we know that there is a good-path $P_\bfx(\bfg_{n-2},\bfg_{n-1})$ in an $\bfx$-type graph  $\bfG_{\bfx}(G)$ for all $\bfx \in \F^{s+1}_2$. Thus, by taking $\bfx = \bfg_{n-2} + \bfy$, the algorithm will execute the reordering function  $\cF_{\bfg_{n-2} + \bfy}(\bfg_{n-2},\bfg_{n-1})$ (Step~\ref{alg:2st:5}).
	By Lemma~\ref{lemma:8}, we obtain two new column vectors $\bfg'_{n-2}$ and $\bfg'_{n-1}$ such that
	\begin{align*}
	\bfg'_{n-2} = &\bfg_{n-2} + \bfg_{n-2} + \bfy = \bfy, \\
	\bfg'_{n-1} = &\bfg_{n-1} + \bfg_{n-2}+\bfy =  \bfw_{n/2-1}+\mathds{1}_{n/2-1}\bfe,
	\end{align*}
	without changing the summation of all other pairs on this path.
	Again, by removing the last row from $G$, we obtain $G'$ such that
	$$\bfg'_{2t} + \bfg'_{2t+1} = \bfv_t, ~~~ t\in [n/2-1]$$
	and $\bfg'_{n-1} = \bfv_{n/2-1}$. Thus, all the recovery sets $R_t, t\in [n/2-1]$ are of the form $R_t = \{2t,2t+1\}$, and  the last recovery set will be $R_{n/2-1} = \{n-1\}$, which concludes this case. In both cases, $G'$ is a double-\HGG with two all-zero columns that will be removed to provide an $FB\bs(2^{s+1}-2,s,2^s)$ code.
\end{IEEEproof}
 
  {For the rest of the paper, we only state that it is possible to obtain the last recovery set from the redundancy columns $\bfg_{n-2}$ and $\bfg_{n-1}$, as it was shown in the proof of Lemma~\ref{lem:22}. }
From the result of Lemma~\ref{lem:22} we deduce the main theorem of this section.
	
\begin{theorem}\label{theo:14}
	An $FB\bs(2^{s+1}-2,s,2^s)$ code exists.
\end{theorem}
 
 \section{A Construction of $FB\bs(2^s-1,s, \lfloor \frac{5}{6} \cdot 2^{s-1}  \rfloor  - s)$ Codes}\label{sec:cons5}
 In this section we show how to improve our main result, i.e., we show a construction of  $FB\bs(2^s-1,s,  \lfloor \frac{5}{6} \cdot 2^{s-1}  \rfloor - s)$ codes. 
Let $M$ be a request denoted by $$M = [\bfv_0, \bfv_1, \dots,\bfv_{  \lfloor \frac{5}{6}\cdot 2^{s-1}  \rfloor - s}].$$
 Remember that $\bfe = (0,0,\dots,0,1) \in \F^s_2$, and $n=2^s$. The initial state of the matrix 
 $$G   = [\bfg_0,\bfg_1,\dots,\bfg_{n-1}]$$ 
 will satisfy
 \begin{align*}
 &\bfg_{2t} + \bfg_{2t+1} = \bfe  , & t\in [n/2].
 \end{align*}
  Remember that for all $\bfx\in\F^s_2$, the graph $\bfG_{\bfx}(G)$ has a partition of $\ell\geq 1$ disjoint simple cycles, that will be denoted by $\bfC_{\bfx}(G) = \{C_i\}^{\ell-1}_{i=0}$ (Definition~\ref{def:5}). Fix $\tau\in [n/2]$. 
The first ingredient in the solution of $FB\bs(2^s-1,s,  \lfloor \frac{5}{6} \cdot 2^{s-1}  \rfloor - s)$ codes will be presented in algorithm \emph{FBSolution2}($\tau, M$), which is presented as Algorithm 5.
 
 \begin{algorithm}[h!]
	 \renewcommand{\thealgorithm}{}
 	\floatname{algorithm}{Algorithm 5}
 	\caption{\emph{FBSolution2}($\tau, M$)}
 	\label{alg:6}
 	\begin{algorithmic}[1]
 		\State  $G^{(0)} \leftarrow G$  \label{alg6:st1}
 		\For{$t=0,\ldots, \tau-1$}\label{alg6:st2}
 		\For {all $2t\leq p,h\leq n-1$ and $t\leq m \leq \tau-1$}\label{alg6:st2.1}
 		\State Swap $\bfg^{(t)}_{p}$ and $\bfg^{(t)}_{2t}$\label{alg6:step5}
 		\State Swap $\bfg^{(t)}_{h}$ and $\bfg^{(t)}_{2t+1}$\label{alg6:step6}
 		\If {$\bfg^{(t)}_{2t} +\bfg^{(t)}_{2t+1} = \bfv_{m}$}\label{alg6:step7}
 		\State Swap between $\bfv_{t}$ and $\bfv_{m}$ in $M$
 		\State Go to Step~\ref{alg6:st2}
 		\EndIf
 		\State $\bfa_m \leftarrow \bfv_m + \bfg^{(t)}_{2t}+\bfg^{(t)}_{2t+1}$\label{alg6:st5.1}
 		\State Let $C_i\in \bfC_{\bfa_m}(G)$ be a cycle s.t. $\{\bfg^{(t)}_{2t},\bfg^{(t)}_{2t+1}\}_\mathbf{p}\in C_i$
 		\If {$\{\bfg^{(t)}_{2\ell},\bfg^{(t)}_{2\ell+1}\}_\mathbf{p}\in C_i$ s.t. $\ell>t$}\label{alg6:st5.2}
 		\State Swap between $\bfv_{t}$ and $\bfv_{m}$ in $M$
 		\State Go to Step~\ref{alg6:st13}
 		\EndIf
 		\EndFor
 		\State  $G^{(t)},M \leftarrow $ \emph{BadCaseCorrection}($G^{(t)},\bfa_m,M$) \label{alg6:st11}
 		\State Go to Step~\ref{alg6:st2.1}
 		 	\State $j \leftarrow$ \emph{FindShortPath}($G^{(t)}, \bfa_m, t,\ell$) \label{alg6:st13}
 		\State $G^{(t+1)}\leftarrow  \cF_{\bfa_t}(\bfg^{(t)}_{2t+1},\bfg^{(t)}_{j})$\label{alg6:step17}
 		\EndFor
 		\State Return $G^{(\tau)}$  \label{alg6:st22}
 	\end{algorithmic}
 \end{algorithm} 
 
 In the internal routine starting on Step~\ref{alg6:st2.1}, on its $t$-th iteration, the algorithm will try to find two column vectors  $\bfg^{(t)}_{p}$ and $\bfg^{(t)}_{h}$, such that $h,p\geq 2t$, and a request $\bfv_{m}$, where $m\geq t$, such that the sum of $\bfg^{(t)}_{p}$ and $\bfg^{(t)}_{h}$ could be updated to $\bfv_m$, without corrupting the sums	$\bfg^{(t+1)}_{2t'} +\bfg^{(t+1)}_{2t'+1}$ such that $t'<t$. Our first task is to prove that if 
 $\tau\leq n/4$ then the algorithm will always find such $\bfg^{(t)}_{p},\bfg^{(t)}_{h}$ and a request $\bfv_{m}$. In this case, the algorithm will provide $2^{s-2}$ (when $\tau = n/4$) requests and will never reach Step~\ref{alg6:st11}. Our second task is to prove that for the case $n/4<\tau\leq \lfloor n/3 \rfloor $, the algorithm may reach Step~\ref{alg6:st11}, however by using the \emph{BadCaseCorrection}($G^{(t)},\bfa_m,M$)  function, which reorders the columns of $G^{(t)}$, the algorithm will succeed to construct $\lfloor\frac{2}{3}\cdot 2^{s-1} \rfloor$ recovery sets of size 2 (when $\tau = \lfloor n/3 \rfloor$). Therefore, we are left to show how to construct additional $\lfloor\frac{1}{6}\cdot 2^{s-1} \rfloor - s + 1$ recovery sets that will be of size $4$ (remember  that $\lfloor x + y \rfloor \leq \lfloor x  \rfloor + \lfloor y \rfloor + 1$).  This part will be handled by the \emph{FBSolution3}($G,\tau, M$) algorithm. 
 {We notice that the \emph{FBSolution3}($G,\tau, M$) algorithm will be invoked only if 
\begin{align}\label{eq:23}
\Big\lfloor\frac{2}{3}\cdot 2^{s-1} \Big\rfloor \leq  \Big\lfloor \frac{5}{6} \cdot 2^{s-1}  \Big\rfloor - s,
\end{align} which holds for $s\geq 7$.}

 \subsection{The $\tau \leq n/4$ Case}
 
 Before proving the correctness of the \emph{FBSolution2}($\tau, M$) algorithm, we start with an important definition.
 
 \begin{definition}
 	On the $t$-th iteration, a path $P_\bfx$ between $\bfg^{(t)}_{p},p\geq 2t$ and $\bfg^{(t)}_{h},h\geq 2t$ will be called a \textbf{short-path} in $\bfG_\bfx(G^{(t)})$, if all the other 
 	pair-type edges $\{\bfg^{(t)}_{2t'},\bfg^{(t)}_{2t'+1}\}_\mathbf{p}$ on $P_\bfx$ satisfy $t'<t$. The short-path $P_{\bfg^{(t)}_{p} +\bfg^{(t)}_{h}}$ is called a trivial short-path.
 \end{definition}

 Our first goal is to show that every $\bfg^{(t)}_{p}, p\geq 2t$ has $n$ different short-paths ending on $n$ columns $\bfg^{(t)}_{h},h\geq 2t$.

\begin{claim}\label{claim:short-path1}
	Fix some $\bfg^{(t)}_{p}$ such that $p\geq 2t$. Then, for each $\bfG_\bfx(G^{(t)})$, there exists $\bfg^{(t)}_{h}, h\geq 2t$,  such that there is a short-path between  $\bfg^{(t)}_{p}$ and $\bfg^{(t)}_{h}$.
\end{claim}

\begin{IEEEproof}
	Given $\bfg^{(t)}_{p}$ such that $p\geq 2t$, its pair $\bfg^{(t)}_{p'}$ also satisfies $p'\geq 2t$. In Claim~\ref{claim:3_2} we proved that for all $\bfx \in \F^s_2$ every pair $\{\bfg_{2m},\bfg_{2m+1}\}_\mathbf{p}$ has a good-path $P_{\bfx}(\bfg_{2m},\bfg_{2m+1})$ in $\bfG_{\bfx}(G)$.  Therefore, by Claim~\ref{claim:3_2}, for all $\bfx \in \F^s_2$, there is a good-path $P_{\bfx}(\bfg^{(t)}_{p},\bfg^{(t)}_{p'})$ in $\bfG_{\bfx}(G^{(t)})$. If for all the edges $\{\bfg^{(t)}_{2m},\bfg^{(t)}_{2m+1}\}_\mathbf{p}$  on $P_{\bfx}(\bfg^{(t)}_{p},\bfg^{(t)}_{p'})$ it holds that $m<t$, then $P_{\bfx}(\bfg^{(t)}_{p},\bfg^{(t)}_{p'})$ is a short-path. Otherwise, there exists an edge $\{\bfg^{(t)}_{2m},\bfg^{(t)}_{2m+1}\}_\mathbf{p}$   on $P_{\bfx}(\bfg^{(t)}_{p},\bfg^{(t)}_{p'})$, such that $m\geq t$, and without loss of generality, we assume that this edge is the closest one to $\bfg^{(t)}_{p}$ on $P_{\bfx}(\bfg^{(t)}_{p},\bfg^{(t)}_{p'})$. Let $h \in\{2m,2m+1\}$ such that the column $\bfg^{(t)}_{h}$ is the closest node between $\bfg^{(t)}_{2m}$ and $\bfg^{(t)}_{2m+1}$ to $\bfg^{(t)}_{p}$ on $P_{\bfx}(\bfg^{(t)}_{p},\bfg^{(t)}_{p'})$. Therefore $h\geq 2t$ and this sub-path is a short-path by definition.
\end{IEEEproof}

 Next, we proceed to prove the correctness of the \emph{FBSolution2}($\tau, M$) algorithm.
 On Step~\ref{alg6:st5.1}, the algorithm will execute $\bfa_m = \bfv_m + \bfg^{(t)}_{p} + \bfg^{(t)}_{h}$. If  $\bfg^{(t)}_{p}$ and $\bfg^{(t)}_{h}$ have a non-trivial short-path between them in $\bfG_{\bfa_m}(G^{(t)})$, then our technique cannot update the sum of  $\bfg^{(t)}_{p}$ and $\bfg^{(t)}_{h}$ to be equal to $\bfv_m$ without changing the sum of a pair $\{\bfg^{(t)}_{2t'},\bfg^{(t)}_{2t'+1}\}_\mathbf{p}$ for some $t'<t$. So our goal is to find columns  $\bfg^{(t)}_{p}$ and $\bfg^{(t)}_{h}$ such that there are no (non-trivial) short-paths between them.
 We state in the following claim that reordering the columns $\bfg^{(t)}_{p}$ for $p\geq 2t$ of $G^{(t)}$ does not affect their short-paths. 
 
 \begin{claim}\label{claim:short-path2}
 	The columns $\bfg^{(t)}_{p}$ and $\bfg^{(t)}_{h}$ (before Steps~\ref{alg6:step5}--\ref{alg6:step6}) have no (non-trivial) short-paths between them, if and only if  the columns $\bfg^{(t)}_{2t}$ and $\bfg^{(t)}_{2t+1}$ (after Steps~\ref{alg6:step5}--\ref{alg6:step6}) have no (non-trivial) short-paths between them. 
 \end{claim}

\begin{IEEEproof}
	We will prove only the first direction, while the second is proved similarly.
	In Claim~\ref{claim:short-path1} we proved that for all $\bfx \in \F^s_2$, there exists $\bfg^{(t)}_{p'}, p'\geq 2t$ such that there is a short-path between  $\bfg^{(t)}_{p}$ and $\bfg^{(t)}_{p'}$. In this claim we assume that every such $p'$ satisfies $p'\neq h$. By definition of the short-path, the edge  $\{\bfg^{(t)}_{2t},\bfg^{(t)}_{2t+1}\}_\mathbf{p}$ is not on any of  these short-paths. Therefore, for all $\bfx \in \F^s_2$, the execution of Step~\ref{alg6:step5} will not affect these short-paths. 
	Similarly, for all $\bfx \in \F^s_2$ the short-paths between $\bfg^{(t)}_{h}$ and $\bfg^{(t)}_{h'}$ will not be affected by the execution of  Step~\ref{alg6:step6} ($h'$ is defined similarly to $h$). Thus, the columns $\bfg^{(t)}_{2t}$ and $\bfg^{(t)}_{2t+1}$ (after Steps~\ref{alg6:step5}--\ref{alg6:step6}) will not have any (non-trivial) short-path between them. 
	
\end{IEEEproof}

 Using Claim~\ref{claim:short-path2}, we can make columns $\bfg^{(t)}_{p}$ and $\bfg^{(t)}_{h}$ to be a pair. This is done by Steps~\ref{alg6:step5} and~\ref{alg6:step6}, i.e., this pair is now  $\{\bfg^{(t)}_{2t},\bfg^{(t)}_{2t+1}\}_\mathbf{p}$. In the next lemma we will use the properties of short-paths to prove the correctness of the algorithm.
 
 \begin{lemma}\label{lemma:good-pathes}
 	On the $t$-th iteration, if there are no (non-trivial) short-paths between $\bfg^{(t)}_{2t}$ and $\bfg^{(t)}_{2t+1}$, then by the end of this iteration it holds that 
 	\begin{align*}
 	\bfg^{(t+1)}_{2t} +\bfg^{(t+1)}_{2t+1}& =\bfv_{m},
 	\end{align*}
 	and all the pair sums	$\bfg^{(t+1)}_{2t'} +\bfg^{(t+1)}_{2t'+1}$ such that $t'<t$ will be unchanged.
 \end{lemma}
 \begin{IEEEproof}
 	If  $\bfg^{(t)}_{2t} +\bfg^{(t)}_{2t+1} = \bfv_{m}$, then due to Step~\ref{alg6:step7} this lemma is correct. Otherwise, we know that there are no (non-trivial) short-paths between $\bfg^{(t)}_{2t}$ and $\bfg^{(t)}_{2t+1}$. Therefore, Step~\ref{alg6:st5.2} will succeed to find $\{\bfg^{(t)}_{2\ell},\bfg^{(t)}_{2\ell+1}\}_\mathbf{p}\in C_i$ such that $\ell>t$. Thus, there is a good-path between $\bfg^{(t)}_{2t+1}$ and one of the nodes $\bfg^{(t)}_{2\ell},\bfg^{(t)}_{2\ell+1}$ (the closest one between them to $\bfg^{(t)}_{2t+1}$), and the index of this node is denoted by $j$ (Step~\ref{alg6:st13}). By executing $\cF_{\bfa_m}( \bfg_{2t+1},\bfg_j)$, the matrix $G^{(t)}$ is updated to a matrix $G^{(t+1)}$ such that only the two following pair summations are correctly changed to
 	\begin{align*}
 	\bfg^{(t+1)}_{2t} +\bfg^{(t+1)}_{2t+1}& = \bfg^{(t)}_{2t} +\bfg^{(t)}_{2t+1}+ \bfa_{m} = \bfv_{m} \\
 	\bfg^{(t+1)}_{2\ell} +\bfg^{(t+1)}_{2\ell+1}& = \bfg^{(t)}_{2\ell} +\bfg^{(t)}_{2\ell+1} + \bfa_m.	
 	\end{align*}
 \end{IEEEproof}

 Next, we will show that	if $t<n/4$ then  the algorithm will find $\bfg^{(t)}_{p}$ and $\bfg^{(t)}_{h}$ with no (non-trivial) short-paths between them.
 
 \begin{lemma}\label{lem:short-pathes}
 	If $t<n/4$, then on the $t$-th iteration the algorithm will find $\bfg^{(t)}_{p}$ and $\bfg^{(t)}_{h}$ with no (non-trivial) short-paths between them.
 \end{lemma}
 \begin{IEEEproof}
 	Fix some $\bfg^{(t)}_{p}$. By Claim~\ref{claim:short-path1}, for each $\bfG_\bfx(G)$, there exists $\bfg^{(t)}_{h}$  such that there is a  short-path between  $\bfg^{(t)}_{p}$ and $\bfg^{(t)}_{h}$. Therefore  $\bfg^{(t)}_{p}$ has $n$ different short-paths. 
 	Since $t<n/4$ or $2t<n/2$, there are at least $n/2+1$ options for choosing $\bfg^{(t)}_{h}$, and each of them has a trivial short-path with $\bfg^{(t)}_{p}$. Therefore, we are left with $n/2-1$ short-paths, and at least $n/2+1$ column vectors $\bfg^{(t)}_{h}$. Thus, there is at least one of them that has no (non-trivial) short-path with $\bfg^{(t)}_{p}$.
 \end{IEEEproof}

 We are now ready to conclude with the following theorem.
 \begin{theorem}
 	For $\tau = n/4$ the algorithm {FBSolution2}($\tau, M$) will construct recovery sums for the first $ 2^{s-2}  $ requests of $M$.
 \end{theorem}
 \begin{IEEEproof}
 	Since $\tau = n/4$, by Lemma~\ref{lem:short-pathes}, on the $t$-th iteration the algorithm \emph{FBSolution2}($\tau, M$) will find $\bfg^{(t)}_{p}$ and $\bfg^{(t)}_{h}$ with no (non-trivial) short-paths between them. Therefore, by Claim~\ref{claim:short-path2}, after executing Steps~\ref{alg6:step5} and~\ref{alg6:step6}, the columns $\bfg^{(t)}_{2t}$ and  $\bfg^{(t)}_{2t+1}$ have no (non-trivial) short-paths between them. 
 	By Lemma~\ref{lemma:good-pathes}, since there are no (non-trivial) short-paths between  $\bfg^{(t)}_{2t}$ and  $\bfg^{(t)}_{2t+1}$, by the end of this iteration it holds that 
 	\begin{align*}
 	\bfg^{(t+1)}_{2t} +\bfg^{(t+1)}_{2t+1}& =\bfv_{m},
 	\end{align*}
 	and all the pair sums	$\bfg^{(t+1)}_{2t'} +\bfg^{(t+1)}_{2t'+1}$ such that $t'<t$ will be unchanged.
 \end{IEEEproof}
 We proceed to the second case, i.e., $\tau \leq \lfloor n/3 \rfloor$.

 \subsection{The $\tau \leq  \lfloor n/3 \rfloor$ Case} 
 
 If $n/4 <  t \leq \lfloor n/3 \rfloor$, then the algorithm may not be able to find $\bfg^{(t)}_{p}$ and $\bfg^{(t)}_{h}$ with no (non-trivial) short-paths between them. However, at least one of these pairs will have at most one (non-trivial) short-path between them. 
 Therefore, we may not be able to make some requests on the $t$-th iteration and reach Step~\ref{alg6:st11}.
 However, if we are left to handle more than one kind of request in $M$, we will succeed in this iteration.
For that, we present the \emph{BadCaseCorrection}($G^{(t)},\bfa_t,M$)  function.
 \begin{algorithm}[h!]
 	\renewcommand{\thealgorithm}{}
 	\floatname{algorithm}{}
 	\caption{\emph{BadCaseCorrection}($G^{(t)},\bfa_t,M$) }
 	\label{alg:7}
 	\begin{algorithmic}[1]
 		\State Let $C_i\in \bfC_{\bfa_t}(G)$ be a cycle s.t. $\{\bfg^{(t)}_{2t},\bfg^{(t)}_{2t+1}\}_\mathbf{p}\in C_i$\label{alg7:st1}
 		\State Find $\{\bfg^{(t)}_{2t'},\bfg^{(t)}_{2t'+1}\}_\mathbf{p}\in C_i$ s.t. $\bfg^{(t)}_{2t'}+\bfg^{(t)}_{2t'+1} \neq \bfv_t$ \label{alg7:st2}
 		 	\State $j \leftarrow$ \emph{FindShortPath}($G^{(t)}, \bfa_t, t,t'$) \label{alg7:st13}
 		\State $G^{(t)}\leftarrow  \cF_{\bfa_t}(\bfg^{(t)}_{2t+1},\bfg^{(t)}_{j})$\label{alg7:st11}
 		\State Swap $\bfg^{(t)}_{2t'}$ and $\bfg^{(t)}_{2t}$\label{alg7:st7}
 		\State Swap $\bfg^{(t)}_{2t'+1}$ and $\bfg^{(t)}_{2t+1}$\label{alg7:st8}
 		\State Swap $\bfv_t$ and $\bfv_{t'}$ in $M$
 		\State Return $G^{(t)},M$
 	\end{algorithmic}
 \end{algorithm}

 \begin{lemma}\label{lemma:hard}
 	On the $t$-th iteration such that $t<\lfloor n/3 \rfloor$, there exist $p,h\geq 2t$ such that there is at most one (non-trivial) short-path between $\bfg^{(t)}_{p}$ and $\bfg^{(t)}_{h}$, and by the end of this iteration it holds that 
 	\begin{align*}
 	\bfg^{(t+1)}_{2t} +\bfg^{(t+1)}_{2t+1}& =\bfv_{m},
 	\end{align*}
	and $t$ recovery sums are satisfied.
 \end{lemma}
 
 \begin{IEEEproof}
 	Fix some $\bfg^{(t)}_{p}$. We already proved that  $\bfg^{(t)}_{p}$ has $n$ different short-paths. Since $t<\lfloor n/3 \rfloor$ or $2t<\lfloor 2n/3 \rfloor$, there are at least $\lceil n/3 \rceil +1 $ options for choosing $\bfg^{(t)}_{h}$, and each one of them has a trivial short-path with $\bfg^{(t)}_{p}$. Therefore, we left with $\lfloor 2n/3 \rfloor -1$ short-paths and $\lceil n/3 \rceil +1$ column vectors $\bfg^{(t)}_{h}$. Therefore, in the worst case, there is  a column vector $\bfg^{(t)}_{h}$ such that there is exactly one (non-trivial) short-path $P_\bfx$ between $\bfg^{(t)}_{h}$ and $\bfg^{(t)}_{p}$. 
 	If $\bfx \neq \bfg^{(t)}_{p} + \bfg^{(t)}_{h} + \bfv_t$, Step~\ref{alg6:st5.2} will succeed, and we can construct a recovery set of size $2$ for $\bfv_t$ on the $t$-th iteration of the \emph{FBSolution2}($\tau, M$) algorithm, as proved in Lemma~\ref{lemma:good-pathes}. If $\bfx = \bfg^{(t)}_{p} + \bfg^{(t)}_{h} + \bfv_t$, we cannot obtain $\bfv_t$ on the $t$-th iteration. So from now we assume that  $\bfx = \bfg^{(t)}_{p} + \bfg^{(t)}_{h} + \bfv_t$.
 	
 	Assume that we have at least two different requests $\bfv_j$ in $M$, and let $\bfv_m\neq \bfv_t$, for $m\geq t$. 
 	We prove that there does not exist a short-path $P_\bfy$ such that $\bfy = \bfg^{(t)}_{p} + \bfg^{(t)}_{h} + \bfv_m$. Therefore, Step~\ref{alg6:st5.2} will succeed to find $\{\bfg^{(t)}_{2\ell},\bfg^{(t)}_{2\ell+1}\}_\mathbf{p}\in C_i$ such that $\ell>t$.  As proved in Lemma~\ref{lemma:good-pathes} the algorithm \emph{FBSolution2}($\tau, M$) will construct a recovery set of size $2$ for $\bfv_m$ on the $t$-th iteration.
 	
 	We are left with considering the case where all the requests $\bfv_m$ in $M$ for $m\geq t$ are identical. Also assume that $\bfg^{(t)}_{2t} +\bfg^{(t)}_{2t+1} \neq  \bfv_{t}$ which means that Step~\ref{alg6:step7} has failed (otherwise this iteration will succeed). In this case the algorithm will use its \emph{BadCaseCorrection}($G^{(t)},\bfa_m,M$)  function.
 	Let $C_i\in \bfC_{\bfa_m}(G)$ be a cycle such that $\{\bfg^{(t)}_{2t},\bfg^{(t)}_{2t+1}\}_\mathbf{p}\in C_i$ (Step~\ref{alg7:st1}).
 	First assume that for every pair-type edge $\{\bfg^{(t)}_{2t'},\bfg^{(t)}_{2t'+1}\}_\mathbf{p}\in C_i$ such that $t'<t$ it holds that
 	\begin{align*}
 	\bfg^{(t)}_{2t'} + \bfg^{(t)}_{2t'+1}=  \bfv_t.
 	\end{align*}
 	In this case, it is easy to verify that it must be a cycle of length $4$, i.e., 
 	\begin{align*}
 	 \bfg^{(t)}_{2t} + \bfg^{(t)}_{2t+1}=	\bfg^{(t)}_{2t'} + \bfg^{(t)}_{2t'+1}=  \bfv_t,
 	\end{align*}
 	which is a contradiction to our assumption.
 	Otherwise, we can assume that there is  an edge $\{\bfg^{(t)}_{2t'},\bfg^{(t)}_{2t'+1}\}_\mathbf{p}\in C_i,t'<t$ such that  
 	\begin{align*}
 	\bfg^{(t)}_{2t'} + \bfg^{(t)}_{2t'+1}= \bfv_{t'} \neq \bfv_t.
 	\end{align*}
 	By executing Step~\ref{alg7:st11}, $\bfg^{(t)}_{2t} + \bfg^{(t)}_{2t+1} $ will be updated to $\bfv_{t}$, which corrupts the sum of the pair $\{\bfg^{(t)}_{2t'},\bfg^{(t)}_{2t'+1}\}_\mathbf{p}$ such that
 	$$\bfg^{(t)}_{2t'} + \bfg^{(t)}_{2t'+1} = \bfv_{t'} + \bfa_t  \neq \bfv_{t'}.$$ 
 	After executing Steps~\ref{alg7:st7}--\ref{alg7:st8}, we obtain two kinds of requests in $M$ that are left to deal with, while still having $t-1$ valid recovery sets. In this case, the algorithm will return to Step~\ref{alg6:st2.1}. Since now we have two kinds of requests, as already proved, the algorithm will be able to construct a recovery set for either $\bfv_t$ or $\bfv_{t'}$ on the $t$-th iteration.
 	
 \end{IEEEproof}

 The next theorem follows directly from Lemma~\ref{lemma:hard}.
 \begin{theorem}~\label{theo:hard}
 	If $\tau\leq \lfloor n/3 \rfloor$, then the algorithm {FBSolution2}($\tau, M$) will construct recovery sums for the first $\lfloor\frac{2}{3}\cdot 2^{s-1} \rfloor$ requests of $M$.
 \end{theorem}
 
Due to Theorem~\ref{theo:hard}, we proved that the algorithm \emph{FBSolution2}($\tau, M$) provides an alternative construction for $FB\bs(2^s-1,s,\lfloor\frac{2}{3}\cdot 2^{s-1} \rfloor)$ codes. However, the algorithm \emph{FBSolution2}($\tau, M$) is better than the algorithm \emph{FBSolution}($\tau, M$) since by using the algorithm \emph{FBSolution2}($\tau, M$), all the recovery sets are of length $2$ (and not $4$). Therefore, we are left with $\lceil n/3 \rceil $ unused column vectors.
 {If~\eqref{eq:23} holds}, we will put $\lceil n/3 \rceil -2 $ of these columns, except the redundant columns $\bfg_{n-2},\bfg_{n-3}$, as the first columns of the \HGG $G$, and similarly, the left requests of $M$ that have no recovery sets yet are placed first. 
In the next section, we show how to obtain $\lfloor\frac{1}{6}\cdot 2^{s-1} \rfloor - s+1$ more recovery sets of size (at most) $4$ from these   $\lceil n/3 \rceil $ unused columns of $G$.

 \subsection{Constructing Recovery Sets of Size $4$}\label{sec:4_4}
 {According to the previous results as stated in Theorem~\ref{theo:0}$(c)$, and due to~\eqref{eq:23}, we can assume that $s\geq 7$.}
 The \emph{FBSolution}($\tau, M$) algorithm uses the initialized \HGG $G^{(0)}$ that satisfies~\eqref{eq:1}. In fact, we can construct a similar algorithm that is initialized by any arbitrary \HGG $G$. Let $\tau =  \lfloor \frac{1}{6}\cdot 2^{s-1} \rfloor -s $. The value of $\tau$ represents the number of requests that will be handled. Note that 
\begin{align*}
4\tau &= 4 \Big\lfloor \frac{1}{6}\cdot 2^{s-1} \Big\rfloor -4s  \\
&= 4 \Big\lfloor \frac{1}{12}\cdot 2^{s} \Big\rfloor -4s  \leq  \Big\lceil \frac{1}{3}\cdot 2^{s} \Big\rceil -2,
\end{align*}
for $s\geq 7$, which is the number of unused columns in $G$.
Our goal is to use either $2$ or $4$ columns of $G$ for every recovery set. In other words, every $\bfv_t$ will be equal to either  $\bfg^{(t)}_{4t} + \bfg^{(t)}_{4t+1}$ or  $\bfg^{(t)}_{4t} + \bfg^{(t)}_{4t+1} +  \bfg^{(t)}_{4t+2} + \bfg^{(t)}_{4t+4}$. To show this property we will prove that in every step of the algorithm, we have to have at least $2(s+1)$ unused (or redundant) columns of $G$. 

We start with the next definition, which is based on the fact that every $s+1$ vectors  in $\F^s_2$ have a subset of $h \leq s+1$ linearly dependent vectors.

\begin{definition}\label{def:20}
	Given an \HGG $G^{(t)}$, denote the set $\cS^{(t)}_h \subseteq [s+1]$ of size $h\leq s+1$, such that
	\begin{align}\label{eq:15}
	\sum_{i \in \cS^{(t)}_h}\Big( \bfg^{(t)}_{4t+2i} + \bfg^{(t)}_{4t+2i+1}\Big)= {\bf{0}}_s.
	\end{align}
	 Denote the  {Reorder}($G^{(t)}$) procedure that swaps arbitrarily between the columns of $G^{(t)}$ presented in~\eqref{eq:15} and the columns indexed by $\{ 4t\leq m \leq 4t+ 2h-1\}$ in $G^{(t)}$ and returns the reordered matrix and and $h$ as an output.
\end{definition}

By using the \emph{Reorder}($G^{(t)}$) procedure which is defined in Definition~\ref{def:20}, we can assume that  
\begin{align}\label{eq:20}
\sum_{ i \in [h]}\Big( \bfg^{(t)}_{4t+2i} + \bfg^{(t)}_{4t+2i+1}\Big)= {\bf{0}}_s.
\end{align}
We are now ready to show the \emph{FBSolution3}($G,\tau, M$) algorithm, which is presented as Algorithm 6.

 \begin{algorithm}[h!]
 \renewcommand{\thealgorithm}{}
 	\floatname{algorithm}{Algorithm 6}
 	\caption{\emph{FBSolution3}($G,\tau, M$)}
 	\label{alg:1_2}
 	\begin{algorithmic}[1]
 		\State $G^{(0)} \leftarrow G$
 		\State $\cB^{(0)} = \emptyset$ 
 		\For{$t=0,1,\ldots, \tau-1$}
 		\State $G^{(t)},h \leftarrow $\emph{Reorder}($G^{(t)}$)
 		\State $G^{(t)} \leftarrow $ \emph{FindEquivSums}($G^{(t)},t,h$)
 		\State  $G^{(t+1)}, \cB^{(t+1)} \leftarrow $ \emph{FindGoodOrBadRequest}($G^{(t)},2t, \bfv_t$)
 		\EndFor
 		\State Return $G^{(\tau)}$ and $\cB^{(\tau)}$
 	\end{algorithmic}
 \end{algorithm}

Note that since 
\begin{align*}
\Big\lceil \frac{1}{3}\cdot 2^{s} \Big\rceil - 2  - 4\tau &=   \Big\lceil \frac{1}{3}\cdot 2^{s} \Big\rceil  - 4\Big\lfloor \frac{1}{12}\cdot 2^{s} \Big\rfloor + 4s  -2\\ 
&\geq 4s  -2 \geq 2(s+1),
\end{align*}
for $s\geq 7$, the $2h \leq 2(s+1)$ column vectors of $G^{(t)}$ presented in~\eqref{eq:20} are unused on the $t$-th iteration. By using these $2h$ unused columns, the function \emph{FindEquivSums}($G,t,h$) will be able to reorder the columns of $G^{(t)}$ such that 
 \begin{align}\label{eq:3}
 \bfg^{(t)}_{4t} + \bfg^{(t)}_{4t+1} =  \bfg^{(t)}_{4t+2} + \bfg^{(t)}_{4t+3},
 \end{align}
without changing the previous valid recovery sums. Then, the function \emph{FindGoodOrBadRequest}($G^{(t)},2t, \bfv_t$) will update the sum $ \bfg^{(t)}_{4t} + \bfg^{(t)}_{4t+1} $ to either $\bfv_t$ or  $\bfg^{(t)}_{4t} + \bfg^{(t)}_{4t+1} + \bfv_t$, again, without changing all the previous valid recovery sums. In the latter case, we are able to construct a recovery set $R_t = \{4t,4t+1,4t+2,4t+3\}$ of size $4$ due to~\eqref{eq:3}.

The \emph{FindEquivSums}($G,t,h$) algorithm is presented next. 

 \begin{algorithm}[h!]
 	\renewcommand{\thealgorithm}{}
 	\floatname{algorithm}{}
	\caption{\emph{FindEquivSums}($G,t,h$).}
	\label{alg:9}
	\begin{algorithmic}[1]
		\For {$i = 1,\ldots,h-1 $}
		\State $\bfx_i \leftarrow \bfg_{4t+2i} + \bfg_{4t + 2i+1}$
		\If{$  \bfg_{4t} + \bfg_{4t+1} \neq \bfx_i   $}\label{alg9:st4}
		\State  $G \leftarrow $ \emph{FindGoodOrBadRequest}($G,2t, \bfx_i $)\label{alg9:st5}
		\EndIf
		\State $\bfx_i \leftarrow \bfg_{4t+2i} + \bfg_{4t + 2i+1}$
		\If{$  \bfg_{4t} + \bfg_{4t+1} = \bfx_i   $}\label{alg9:st6}
		 \State Swap $\bfg_{4t+2}$ and $\bfg_{4t+2i}$
		\State Swap $\bfg_{4t+3}$ and $\bfg_{4t+2i+1}$
		\State Return $G$
		\EndIf
		\EndFor
	\end{algorithmic}
\end{algorithm}

The proof of the correctness of the function \emph{FindEquivSums}($G,t,h$) is shown in the following theorem.
\begin{theorem}
	 If there are at least $2(s+1)$ unused columns in $G$, then there is a function {FindEquivSums}($G,t,h$) that can reorder the columns of $G$ such that 
	  \begin{align*}
	 \bfg_{4t} + \bfg_{4t+1} =  \bfg_{4t+2} + \bfg_{4t+3},
	 \end{align*}
	  without corrupting the previous valid recovery sums. 
\end{theorem} 

\begin{IEEEproof}
	As explained before, the $2h \leq 2(s+1)$ column vectors of $G$ presented in~\eqref{eq:20} are unused. Therefore, the algorithm \emph{FindEquivSums}($G,t,h$) will not corrupt the previous valid recovery sums. Our goal is to prove that the algorithm will succeed on Step~\ref{alg9:st6}. On the $i$-th iteration, by executing Step~\ref{alg9:st5}, the sum of  $\bfg_{4t} + \bfg_{4t+1}$ will be either $ \bfg_{4t} + \bfg_{4t+1} + \bfx_i$ or  $\bfx_i$, without changing other sums except of the redundancy sum $\bfg_{n-2} + \bfg_{n-1}$. If $\bfg_{4t} + \bfg_{4t+1} = \bfx_i$, Step~\ref{alg9:st6} will succeed. Otherwise, we can assume that on the $i$-th iteration, the algorithm obtains $ \bfg_{4t} + \bfg_{4t+1} + \bfx_i$ on Step~\ref{alg9:st5}. 
	Denote by $\bfx_0$ the sum of $ \bfg_{4t} + \bfg_{4t+1}$ at the beginning of the algorithm. Therefore, at the end of the $i$-th iteration, we obtain
	\begin{align*}
	 \bfg_{4t} + \bfg_{4t+1} = \sum^{i}_{  j=0} \bfx_j.
	\end{align*}	
	By~\eqref{eq:20},
	\begin{align*}
	\sum^{h-2}_{  j=0} \bfx_j= \bfx_{h-1}.
	\end{align*}
	Therefore, on the last iteration, i.e., when $i=h-1$,
	\begin{align*}
	 \bfg_{4t} + \bfg_{4t+1} = \sum^{h-2}_{  j=0} \bfx_j = \bfx_{h-1}.
	\end{align*}
	 Thus, Step~\ref{alg9:st4} will fail and Step~\ref{alg9:st6} will succeed, concluding the proof.
\end{IEEEproof}

 We are ready to show the main result of this section. 
 
 \begin{lemma}\label{lemma:10}
 	The  {FBSolution3}($G,\tau, M$) algorithm constructs $\tau$ valid recovery sets $R_t$ for $\bfv_t$, without corrupting previous recovery sums.
 \end{lemma}

\begin{IEEEproof}
By Lemma~\ref{lemma:10} the \emph{FindEquivSums}($G^{(t)},t, M$) algorithm will output $G^{(t)}$ such that $\bfg^{(t)}_{4t} + \bfg^{(t)}_{4t+1}= \bfg^{(t)}_{4t+2} + \bfg^{(t)}_{4t+2}$, and all previous sums of requests are valid.
Before the execution of the \emph{FindGoodOrBadRequest}($G^{(t)},2t, \bfv_t$) function we denote the sums $\bfg_{4t} + \bfg_{4t+1}$ and  $ \bfg_{4t+2} + \bfg_{4t+2}$ by $\bfy_t$. 
The algorithm \emph{FindGoodOrBadRequest}($G^{(t)},2t, \bfv_t$) will update only the sum  $\bfg^{(t)}_{4t} + \bfg^{(t)}_{4t+1}$ to either $\bfv_t$ or  $\bfv_t + \bfy_t$  and the sum of the last pair $\bfg_{n-2} + \bfg_{n-3}$ which is redundant, due to Lemma~\ref{lemma:11}. The case $\bfg^{(t)}_{4t} + \bfg^{(t)}_{4t+1} = \bfv_t$ is called a good case and  we assume that all such $t$'s are inserted in a set $\cG$. These pairs will be recovered by the recovery sets $R_t = \{4t,4t+1\}$.
The case $\bfg^{(t)}_{4t} + \bfg^{(t)}_{4t+1} = \bfv_t + \bfy_t$ is called a bad case and all such $t$'s  are assumed to  be inserted in a set $\cB$. By~\eqref{eq:3} for every $t$  such that $\bfg^{(t)}_{4t} + \bfg^{(t)}_{4t+1} = \bfv_t + \bfy_t$ we have that $\bfg^{(t)}_{4t+2} + \bfg^{(t)}_{4t+3} = \bfy_t$. Thus, in these cases, the requests will have the recovery sets $R_t = \{4t,4t+1,4t+2,4t+3\}$.
\end{IEEEproof}

In Section~\ref{sec:cons2}, we showed a technique to obtain another recovery set from the redundancy pair $\bfg_{n-2}$ and $\bfg_{n-1}$. Using this technique, we are able to construct  $\lfloor\frac{1}{6}\cdot 2^{s-1} \rfloor - s + 1$ valid recovery sets.
By combining the three cases above, the following theorem is deduced immediately.
\begin{theorem}\label{theo:15}
	An $FB\bs(2^s-1,s,  \lfloor \frac{5}{6}\cdot 2^{s-1}  \rfloor  - s)$ code exists.
\end{theorem}

\section{A Construction of $B\bs(2^{s}-1,s,2^{s-1})$  Codes}\label{sec:extension1}

Wang et al.~\cite{Wang} showed a construction for $B\bs(2^{s}-1,s,2^{s-1})$ codes, which is optimal, using a recursive decoding algorithm.   In this section, we show how to achieve this result with the simpler, non-recursive  {decoding} algorithm. Our solution solves even a more general case in which the requests $\bfv_j$'s satisfy some constraint that will be described later in this section. The idea of this algorithm is similar to the one of the \emph{FBSolution}($\tau, M$) algorithm. First, we slightly change the definition of a Hadamard solution as presented in Definition~\ref{def:6.1} to be the following one.

\begin{definition}\label{def:6.3}
	Let $ M  = [\bfv_0, \bfv_1, \dots,\bfv_{n/2-1} ]$ be a request of order $s\times n/2$, where $n=2^s$. The matrix $M$ has a \textbf{Hadamard solution} if there  exists an \HGG  	$G = [\bfg_0,\bfg_1,\dots,\bfg_{n-1}]$ of order $s\times n$ such  that for all  $i\in [n/2-1]$,
	\begin{align*}
	&\bfv_i = \bfg_{ 2i } + \bfg_{2i+1},
	\end{align*}
	and for $i=n/2-1$ either $\bfv_i =  \bfg_{ n-2 } + \bfg_{n-1}$, or $\bfv_i = \bfg_{n-2}$, or $\bfv_i = \bfg_{n-1}$.
	In this case, we say that $G$ is a Hadamard solution for $M$.
\end{definition}

Let $G$ be an \HG. Let $\mathbf{G}$ be the set of all matrices $G'$ generated  by elementary row operations on $G$. The following claim proves that elementary row operations on \HGs only reorder their column vectors. 

\begin{claim}\label{claim:gauss}
	Every $G \in \mathbf{G}$ is an \HG.
\end{claim}

\begin{IEEEproof}
We will only prove that adding a row in $G$ to any other row, generates an \HG. By proving that, it can be inductively proved that doing several such operations will again yield an \HG.

Without loss of generality, we assume that we add the $i$-th row, for some $0<i\leq s-1$, to the $0$-th row of $G$ and generate a new matrix $G'$.
Assume to the contrary that  $G'$ is not an \HG. Thus, there are two distinct indices $\ell,m \in [n]$ such that $\bfg'_\ell = \bfg'_m$. Therefore, by definition of elementary row operations,  $G$ satisfies  $\bfg_\ell = \bfg_m$, which is a contradiction. 
\end{IEEEproof}	

Let $M$ be a request denoted by $$M = [\bfv_0, \bfv_1, \dots,\bfv_{ 2^{s-1}-1}].$$
Let $\mathbf{M}$ be the set of all matrices $M'$ generated  by elementary row operations on request matrix $M$. We now present Lemma~\ref{lem:gauss}. Its proof follows directly from Claim~\ref{claim:gauss}.

\begin{lemma}\label{lem:gauss}
	If there is an $M\in\mathbf{M}$ such that there is a Hadamard solution for $M$, then there is a Hadamard solution  for all $M'\in \mathbf{M}$.
\end{lemma}

\begin{IEEEproof}
	 Let $M\in\mathbf{M}$ and let $G$ be a Hadamard solution for $M$. Let $P$ be a set of elementary row operations, generating $M'$ from $M$. By Claim~\ref{claim:gauss}, executing elementary row operations $P$ on $G$ generates   an \HG, $G'$. Since we applied the same elementary row operations $P$ on both $M$ and $G$, it is deduced that $G'$ is a Hadamard solution for $M'$.
	
\end{IEEEproof}


 The constraint mentioned above is as follows. Given $\mathbf{M}$, we demand that there is a request $M'\in\mathbf{M}$ having the $0$-th row to be a vector of ones. Using Lemma~\ref{lem:gauss}, our algorithm will handle any request $M'$ such that $M'\in\mathbf{M}$ and $\mathbf{M}$ holds this constraint.
Note that if each request vector $\bfv_j\in\F^s_2$ is a unit vector, then by summing up all of its rows to the $0$-th one, it holds that there exists such a matrix in $\mathbf{M}$ holds the constraint. Moreover, if every request vector is of odd Hamming weight, our algorithm will still find a solution.  Therefore, from now on, we assume that the $0$-th row of the request matrix $M$ is a vector of ones.

Remember that $\bfe = (0,0,\dots,0,1) \in \F^s_2$. The initial state of the matrix 
$$G =  [\bfg_0,\bfg_1,\dots,\bfg_{n-1}]$$ 
will satisfy
\begin{align*}
&\bfg_{2t} + \bfg_{2t+1} = \bfe  , & t\in [n/2].
\end{align*}
Given $\bfg\in \F^s_2$, let $f(\bfg)$ be the bit value in the $0$-th position in $\bfg$. Remember that for all $\bfx\in\F^s_2$, the graph $\bfG_{\bfx}(G)$ has a partition to $\ell\geq 1$ disjoint simple cycles, that will be denoted by $\bfC_{\bfx}(G) = \{C_i\}^{\ell-1}_{i=0}$ (Definition~\ref{def:5}). 
Let $\tau = 2^{s-1}$. 
We are now ready to show the following algorithm.

\begin{algorithm}[h!]
	  \renewcommand{\thealgorithm}{}
	 \floatname{algorithm}{Algorithm 7}
	\caption{\emph{BSolution}($\tau, M$)}
	\label{alg:5}
	\begin{algorithmic}[1]
		\State  $G^{(0)} \leftarrow G$  \label{alg5:st1}
		\For{$t=0,\ldots, \tau-2$}\label{alg5:st2}
		\If {$f(\bfg^{(t)}_{2t}+\bfg^{(t)}_{2t+1})=1$}\label{alg5:step3}
			\State Find $\bfg^{(t)}_{p},\bfg^{(t)}_{h}$ s.t. $p,h\geq2t$ and $f(\bfg^{(t)}_{p}+\bfg^{(t)}_{h})=0$
			\State Swap $\bfg^{(t)}_{p}$ and $\bfg^{(t)}_{2t}$\label{alg5:step5}
			\State Swap $\bfg^{(t)}_{h}$ and $\bfg^{(t)}_{2t+1}$\label{alg5:step6}
		\EndIf
		\State $\bfa_t \leftarrow \bfv_t + \bfg^{(t)}_{2t}+\bfg^{(t)}_{2t+1}$\label{alg5:st5}
		\State Let $C_i\in \bfC_{\bfa_t}(G^{(t)})$ be a cycle s.t. $\{\bfg^{(t)}_{2t},\bfg^{(t)}_{2t+1}\}_\mathbf{p}\in C_i$
		\State Find $\{\bfg^{(t)}_{2m},\bfg^{(t)}_{2m+1}\}_\mathbf{p}\in C_i$ s.t. $m>t$\label{alg5:st21_1}
		 	\State $j \leftarrow$ \emph{FindShortPath}($G^{(t)}, \bfa_t, t,m$) \label{alg5:st16}
		\State $G^{(t+1)}\leftarrow  \cF_{\bfa_t}(\bfg^{(t)}_{2t},\bfg^{(t)}_{j})$\label{alg5:step17}
		\EndFor
		\If{$\bfg^{(\tau-1)}_{n-2}+ \bfg^{(\tau-1)}_{n-1}\neq \bfv_{n/2-1}$ and $\bfg^{(\tau-1)}_{n-2}\neq \bfv_{n/2-1}$}\label{alg5:step18}
		\State $G^{(\tau)} \leftarrow  \cF_{\bfg^{(\tau-1)}_{n-2} + \bfv_{n/2-1}}(\bfg^{(\tau-1)}_{n-2},\bfg^{(\tau-1)}_{n-1})$\label{alg5:step19}
		\EndIf
		\State Return $G^{(\tau)}$  \label{alg5:st22}
 	\end{algorithmic}
\end{algorithm} 

Our first goal is to prove that on the $t$-th iteration when the \emph{BSolution}($\tau, M$) algorithm reaches Step~\ref{alg5:st5}, it holds that $f(\bfg^{(t)}_{2t}+\bfg^{(t)}_{2t+1})=0$.

\begin{lemma}\label{lemma:cycles1}
	On the $t$-th iteration when the {BSolution}($\tau, M$) algorithm reaches Step~\ref{alg5:st5}, it holds that $f(\bfg^{(t)}_{2t}+\bfg^{(t)}_{2t+1})=0$.
\end{lemma}
\begin{IEEEproof}
	If on the $t$-th iteration in Step~\ref{alg5:step3}, $f(\bfg^{(t)}_{2t}+\bfg^{(t)}_{2t+1})=1$ then the \emph{BSolution}($\tau, M$) algorithm will try to find $p,h\geq2t$ such that $f(\bfg^{(t)}_{p}+\bfg^{(t)}_{h})=0$. Note that since $t\leq \tau-2$, we have $\bfg^{(t)}_{2t},\bfg^{(t)}_{2t+1}$ and at least two more column vectors $\bfg^{(t)}_{2m},\bfg^{(t)}_{2m+1}$ such that $m>t$. By the pigeonhole principle, there exist two indices $p,h\in\{2t,2t+1,2m,2m+1\}$ such that $f(\bfg^{(t)}_{p})=f(\bfg^{(t)}_{h})$. After executing Step~\ref{alg5:step5} and Step~\ref{alg5:step6} we obtain $f(\bfg^{(t)}_{2t}+\bfg^{(t)}_{2t+1})=0$.
\end{IEEEproof}

Our next goal is to show that in Step~\ref{alg5:st21_1} on the $t$-th iteration, the \emph{BSolution}($\tau, M$) algorithm will find $\{\bfg^{(t)}_{2m},\bfg^{(t)}_{2m+1}\}_\mathbf{p}\in C_i$ such that $m>t$. We start with the following claim.

\begin{claim}\label{claim:cycles}
	Given a graph $\bfG_{\bfx}(G) $ and its partition to cycles $\bfC_{\bfx}(G) = \{C_i\}^{p-1}_{i=0}$, for all $C_i\in \bfC_{\bfx}(G) $ it holds that 
	\begin{align*}
	\sum_{ \{\bfg_{2m},\bfg_{2m+1}\}_\mathbf{p} \in C_i }\Big( \bfg_{2m}+\bfg_{2m+1} \Big) = \frac{1}{2}|C_i|\bfx,
	\end{align*}
	where the operations are over the binary field.
\end{claim}
\begin{IEEEproof}
	Assume that $C_i$ is of length $2\ell$, and its cycle representation is given as follows 
	\begin{align*}
	C_i  = \bfg_{s_0} - \bfg_{s_1} - \dots - \bfg_{s_{2\ell-1}} - \bfg_{s_{2\ell}}-\bfg_{s_0},
	\end{align*}
	where both of the edges $\{\bfg_{s_0},\bfg_{s_1}\}_\mathbf{p}, \{\bfg_{s_{2\ell-1}},\bfg_{s_{2\ell}}\}_\mathbf{p}$ are pair type edges.
	By Claim~\ref{claim:3}\eqref{claim:3_4}, for all odd $t\in [2\ell]$ it holds that $\bfg_{s_t}=\bfg_{s_{t+1}} + \bfx$.
	Thus,  {by summing only the sums of the nodes of the pair-type edges in $C_i$ we obtain}
	\begin{align*}
	&	\sum_{ \{\bfg_{2m},\bfg_{2m+1}\}_\mathbf{p} \in C_i }\Big( \bfg_{2m}+\bfg_{2m+1} \Big)\\
	& =  \sum_{ t \in [2\ell] } \bfg_{s_i}   =  \sum_{ t \in [2\ell], t \textrm{ is odd} }\Big(\bfg_{s_t} + \bfg_{s_t} + \bfx\Big)  \\
	& = \ell\bfx.
	\end{align*}
	
\end{IEEEproof}

Now we are ready to prove the following lemma.

\begin{lemma}\label{lem:cycles2}
	On the $t$-th iteration when the {BSolution}($\tau, M$) algorithm reaches Step~\ref{alg5:st21_1}, it will find $\{\bfg^{(t)}_{2m},\bfg^{(t)}_{2m+1}\}_\mathbf{p}\in C_i$ such that $m>t$.
\end{lemma}

\begin{IEEEproof}
	Remember that we assumed that the bit value in the $0$-th position for all the requests $\bfv_j$ is $1$. Therefore, on the $t$-th iteration, for all $m<t$ it holds that $f(\bfg^{(t)}_{2m}+\bfg^{(t)}_{2m+1})=1$, and by Lemma~\ref{lemma:cycles1}, $f(\bfg^{(t)}_{2t}+\bfg^{(t)}_{2t+1})=0$. Now, assume to the contrary that there are no $m>t$ such that $\{\bfg^{(t)}_{2m},\bfg^{(t)}_{2m+1}\}_\mathbf{p}\in C_i$. 
	Note that 
	\begin{align*}
	f(\bfa_t) &= f(\bfv_t + \bfg^{(t)}_{2t}+\bfg^{(t)}_{2t+1}) \\
	&=f(\bfv_t) + f(\bfg^{(t)}_{2t}+\bfg^{(t)}_{2t+1}) = 1 + 0 =  1  {(\bmod~2).}
	\end{align*}
	By Claim~\ref{claim:cycles}, if $|C_i|=2\ell$ then,
	\begin{align*}
	\sum_{ \{\bfg^{(t)}_{2m},\bfg^{(t)}_{2m+1}\}_\mathbf{p} \in C_i } \Big(\bfg^{(t)}_{2m}+\bfg^{(t)}_{2m+1}\Big) = \ell\bfa_t,
	\end{align*}
	and therefore, 
	\begin{align*}
	\sum_{ \{\bfg^{(t)}_{2m},\bfg^{(t)}_{2m+1}\}_\mathbf{p}\in C_i } f(\bfg^{(t)}_{2m}+\bfg^{(t)}_{2m+1}) = \ell f(\bfa_t) = \ell.
	\end{align*}
	However, since only the edge $\{\bfg^{(t)}_{2t},\bfg^{(t)}_{2t+1}\}_\mathbf{p} \in C_i $ satisfies that $f(\bfg^{(t)}_{2t}+\bfg^{(t)}_{2t+1}) = 0$, it is deduced that
	\begin{align*}
	\sum_{ \{\bfg_{2m},\bfg_{2m+1}\}_\mathbf{p} \in C_i } f(\bfg^{(t)}_{2m}+\bfg^{(t)}_{2m+1})= \ell-1  \not\equiv  {\ell}(\bmod~2),
	\end{align*}
	which violates Claim~\ref{claim:cycles}.
\end{IEEEproof}

We are ready to show the main theorem of this section.
\begin{theorem}
	Given a request matrix $M$ having the $0$-th row to be a vector of ones, the {BSolution}($\tau, M$) algorithm finds a Hadamard solution for $M$. 
\end{theorem}
\begin{IEEEproof}
	First, we will prove that the \emph{BSolution}($\tau, M$) algorithm generates $2^{s-1}-1$ recovery sets for the first $2^{s-1}-1$ requests $\bfv_t$. This is done by Steps~\ref{alg5:st1}--\ref{alg5:step17}. Note that the sums $\bfg^{(t)}_{2m}+\bfg^{(t)}_{2m+1}$ for all $m<t$,  might be changed only after Step~\ref{alg5:st21_1}. We will show that these sums will not be changed and the sum  $\bfg^{(t)}_{2t}+\bfg^{(t)}_{2t+1}$ will be equal to $\bfv_t$ at the end of the $t$-th iteration.
	By Lemma~\ref{lem:cycles2}, when the \emph{BSolution}($\tau, M$) algorithm reaches Step~\ref{alg5:st21_1}, it will find $\{\bfg^{(t)}_{2m},\bfg^{(t)}_{2m+1}\}_\mathbf{p}\in C_i$ such that $m>t$. Thus, there is a good-path between $\bfg_{2t+1}$ and one of the nodes $\bfg_{2m},\bfg_{2m+1}$ (the closest one between them to $\bfg_{2t+1}$), and the index of this node is denoted by $j$ (Step~\ref{alg5:st16}). Due to Lemma~\ref{lemma:8}, by executing $\cF_{\bfa_t}( \bfg_{2t+1},\bfg_j)$, the matrix $G^{(t)}$ is updated to a matrix $G^{(t+1)}$ such that only the two following pair summations are correctly changed to
	\begin{align*}
	\bfg^{(t+1)}_{2t} +\bfg^{(t+1)}_{2t+1}& = \bfg^{(t)}_{2t} +\bfg^{(t)}_{2t+1}+ \bfa_{t} = \bfv_{t} \\
	\bfg^{(t+1)}_{2m} +\bfg^{(t+1)}_{2m+1}& = \bfg^{(t)}_{2m} +\bfg^{(t)}_{2m+1} + \bfa_t.	
	\end{align*}
	Lastly, Steps~\ref{alg5:step18}--\ref{alg5:step19} handle the last recovery set in a similar way as was done in the proof of Theorem~\ref{theo:16}.
\end{IEEEproof}

\section{Conclusion}\label{sec:conc}
 {In this paper, functional $k$-batch codes and the value $FB(s,k)$ were studied. It was shown that for all $s\geq 6$, $FB(s,\lfloor \frac{5}{6}2^{s-1} \rfloor-s) \leq 2^{s}-1$. In fact, we believe that by using a similar technique, this result can be improved to $\lfloor \frac{7}{8}2^{s-1} \rfloor-s$ requests, but this proof has many cases and thus it is left for future work.
We also showed a family of $FB\bs(2^s +\lceil (3\alpha-2)\cdot2^{s-2}\rceil -1 ,s,\lfloor \alpha\cdot 2^{s-1} \rfloor)$ codes for all $2/3 \leq \alpha \leq 1$. Yet another result in the paper provides an optimal solution for $k=2^s$ which is $FB(s,2^s) = 2^{s+1}-2$. 
While the first and main result of the paper significantly improves upon the best-known construction in the literature, there is still a gap to the conjecture which claims that $FB(s,2^{s-1}) = 2^s-1$. We believe that the conjecture indeed holds true and it can be achieved using Hadamard codes. 
}


\begin{thebibliography}{1}


	
	\bibitem{Asi}
	H. Asi and E. Yaakobi,
	``Nearly optimal constructions of PIR and batch	codes,"
	\emph{In Proceedings IEEE International Symposium on Information Theory},
	pp. 151--155, Aachen, Germany, Jun. 2017.


	
	\bibitem{Buzaglo}
	S. Buzaglo, Y. Cassuto, P. H. Siegel, and E. Yaakobi,
	``Consecutive switch codes,"
	\emph{IEEE Transactions on Information Theory},
	vol. 64, no. 4, pp. 2485--2498,	Apr. 2018.
	
	
	
	\bibitem{Arora}
	S. Arora and B. Barak,
	``Computational complexity -- a modern approach,"
	\emph{Cambridge University Press},
	Cambridge, 2009.
	
	
	\bibitem{Cohen}
	G.D. Cohen, P. Godlewski, and F. Merx,
	``Linear binary code for write-once memories,"
	\emph{IEEE Transactions on Information Theory},
	vol. 32, no. 5, pp. 697--700, Oct. 1986.	
	
	
	\bibitem{Fazeli}
	A. Fazeli, A. Vardy, and E. Yaakobi,
	``PIR with low storage overhead: Coding instead of replication,"
	\emph{arxiv.org/abs/1505.06241}, May 2015.
	
	\bibitem{Godlewski}
	P. Godlewski,
	``WOM-codes construits {\`a} partir des codes de Hamming,"
	\emph{Discrete Mathematics},
	vol. 65, no. 3, pp. 237--243, Jul. 1987.

	\bibitem{Ishai}
	Y. Ishai, E. Kushilevitz, R. Ostrovsky, and A. Sahai,
	``Batch codes and their applications,"
	\emph{In Proceedings 36th Annual ACM Symposium on Theory of Computing},
	pp. 262--271, 2004.



	\bibitem{Rawat}
	A. S. Rawat, Z. Song, A. G. Dimakis, and A. G{\'a}l,
	``Batch codes through dense graphs without short cycles,"
	\emph{IEEE Transactions on Information Theory},
	vol. 62, no. 4, Apr. 2016.
	
	
	
	\bibitem{Rivest}
	R.L. Rivest and A. Shamir,
	``How to reuse a write-once memory,"
	\emph{Information and Control},
	vol. 55, no. 1--3, pp. 1--19, Dec. 1982.


	\bibitem{Sharon}
	E. Sharon and I. Alrod,
	``Coding scheme for optimizing random I/O performance,"
	\emph{Non-Volatile Memories Workshop},	San Diego, Mar. 2013.


	\bibitem{Sun}
	H. Sun and S. A. Jafar,
	``The capacity of private computation,"
	\emph{arxiv.org/abs/1710.11098}, Oct. 2017.	


	\bibitem{Vajha}
	M. Vajha, V. Ramkumar, and P. V. Kumar,
	`` Binary, shortened projective Reed Muller codes for coded private information retrieval,"
	\emph{In Proceedings IEEE International Symposium on Information Theory},
	pp. 2648--2652. Aachen, Germany, Jun. 2017.

	\bibitem{Vardy}
	A. Vardy and E. Yaakobi,
	``Constructions of batch codes with near optimal redundancy,"
	\emph{In Proceedings IEEE International Symposium on Information Theory},
	pp. 1197--1201, 2016.
	
	\bibitem{Wang}
	Z. Wang, H. M. Kiah, and Y. Cassuto,
	``Switch codes: Codes for fully parallel reconstruction" 
	\emph{IEEE Transactions on Information Theory},
	vol. 63, no. 4, pp. 2061--2075, Apr. 2017.

		
	\bibitem{Yaakobi}
	E. Yaakobi, S. Kayser, P. H. Siegel, A. Vardy, and J.K. Wolf,
	``Codes for write-once memories,"
	\emph{IEEE Transactions on Information Theory},
	vol. 58, no. 9, pp. 5985--5999, Sep. 2012.

	\bibitem{Yamawaki}
	A. Yamawaki, H. Kamabe, and S. Lu,
	``Construction of parallel RIO codes using coset coding with Hamming code,"
	\emph{In Proceedings IEEE Information Theory Workshop},
	pp. 239--243, Kaohsiung, Taiwan, Nov. 2017.

	\bibitem{Zhang}
	Y. Zhang, T. Etzion, and E. Yaakobi,
	``Bounds on the length of functional PIR and batch codes,"
	\emph{IEEE Transactions on Information Theory},
	vol. 66, no. 8, pp. 4917--4934, Aug. 2020.




\end{thebibliography}
\end{document}